\documentclass[aps,pre,twocolumn,groupedaddress,amsmath]{revtex4}

\usepackage{graphicx}
\usepackage{psfrag}

\psfrag{Ccal}[B1][B1][1]{${\mathcal{C}}/R$}
\psfrag{Ccal1}[B1][B1][1]{${\mathcal{C}}$}
\psfrag{Vcal}[B1][B1][1]{${\mathcal{V}}$}
\psfrag{Ecal}[B1][B1][1]{${\mathcal{E}}$}
\psfrag{Thetai}[B1][B1][1]{$\theta_i$}

\begin{document}

\title{Forward Flux Sampling-type schemes for simulating rare events: Efficiency analysis}

\author{Rosalind J. Allen}


\affiliation{FOM Institute for Atomic and Molecular Physics, Kruislaan 407, 1098 SJ Amsterdam, The Netherlands}

\author{Daan Frenkel}

\affiliation{FOM Institute for Atomic and Molecular Physics, Kruislaan 407, 1098 SJ Amsterdam, The Netherlands}

\author{Pieter Rein ten Wolde}
 
\email{tenwolde@amolf.nl}
 
\affiliation{FOM Institute for Atomic and Molecular Physics, Kruislaan 407, 1098 SJ Amsterdam, The Netherlands}

\date{\today}

\begin{abstract}
We analyse the efficiency of several simulation methods which we have recently proposed for calculating rate constants for rare events in stochastic dynamical systems, in or out of equilibrium.  We derive analytical expressions for the computational cost of using these methods, and for the statistical error in the final estimate of the rate constant, for a given computational cost. These expressions can be used to determine which method to use for a given problem, to optimize the choice of parameters, and to evaluate the significance of the results obtained. We apply the expressions to the two-dimensional non-equilibrium rare event problem proposed by Maier and Stein. For this problem, our analysis gives accurate quantitative predictions for the computational efficiency of the three methods.
\end{abstract}


\maketitle

\section{Introduction}
Rare events are processes which happen rapidly, yet infrequently. Specialized techniques are required in order to study these events using computer simulation. This is because, in ``brute force'' simulations, the  vast majority of the computational effort is used in  simulating the uninteresting waiting periods between events, so that  observing enough events for reliable statistical analysis is generally impossible. The quantities of interest from the simulation point of view are generally the rate constant for the rare transitions between the initial and final states and the properties of the Transition Path Ensemble (TPE) - the (correctly weighted) collection of transition trajectories. When computing these quantities, it is very important to know the statistical error in the calculated value, and the likely cost of the computation. In this paper, we derive approximate expressions for these quantities, for three rare event simulation methods which we proposed in a recent publication \cite{long1}. These expressions turn out to be surprisingly accurate for simulations of a model rare event problem. Our results allow us to quantify the computational efficiency of the three methods.

The three ``FFS-type'' simulation methods allow the computation of both the rate constant and the transition paths for rare events in equilibrium or non-equilibrium steady-state systems with stochastic dynamics. In all three methods, a series of interfaces are defined between the initial and final states.  The rate constant is given by the flux of trajectories crossing the first interface, multiplied by the probability that these trajectories subsequently reach B. The latter probability is  computed by carrying out a series of ``trial'' runs between successive interfaces; this procedure also generates transition paths, which are chains of connected successful trial runs. The methods differ in the way the trial runs are fired and the transition paths are generated.   In the ``forward flux sampling'' (FFS) method, a collection of points is generated at the first interface and trial runs are used to propagate this collection of points to subsequent interfaces - thus generating many transition paths simultaneously. In the branched growth (BG) method, a single point is generated at the first interface and is used as the starting point for multiple trial runs to the next interface. Each successful trial generates a starting point for multiple trials to the following interface, so that a ``branching tree'' of transition paths is generated. In the Rosenbluth (RB) method, a single starting point is chosen at the first interface, multiple trial runs are carried out, but only one successful trial is used to propagate the path to the next interface - thus unbranched transition paths are generated. In this method, a re-weighting step is needed to ensure correctly weighted transition paths. 

A range of simulation techniques for rare events in soft condensed matter systems are currently available. In  Bennett-Chandler-type methods, the rate constant is obtained via a computation of a free energy barrier \cite{daan}. In Transition Path Sampling (TPS) \cite{tps},  transition trajectories (paths) are  generated by shooting forwards and backwards in time from already existing paths, and are then sampled using a Monte Carlo procedure. The rate constant is obtained via the computation of a time correlation function. Bennett-Chandler-type methods and TPS are suitable for systems with stochastic or deterministic dynamics, but they require knowledge of the steady state phase space density, which means that the system must be in equilibrium. While the FFS-type methods are only suitable for systems with stochastic dynamics, they do not require the phase space density to be known and can therefore be used for non-equilibrium steady states not satisfying detailed balance. To our knowledge, the only other path sampling method that is suitable for non-equilibrium systems is that proposed recently by Crooks and Chandler \cite{crooks}, which adopts a ``TPS''-type methodology, generating new stochastic paths from old paths by changing the random number history.

The origin of the efficiency of the FFS-type methods is that they use a series of interfaces in phase space between the initial and final states to divide up the transition paths into a series of connected ``partial paths''. These partial paths are generated in a ratchet-like manner - {\em{i.e.}} once a particular interface has been reached, the system configuration is stored and is used to initiate trial runs to the next interface.  Many other rare event techniques also use a series of interfaces in phase space. In Transition Interface Sampling (TIS) \cite{vanerp} and Partial Path Transition Interface Sampling (PPTIS) \cite{moroni}, interfaces are used to facilitate the generation of transition paths by a TPS-like procedure. In Milestoning \cite{faradjian}, trajectories are generated between interfaces assuming a steady-state distribution at each interface, while string methods \cite{e2002,e2005} use a series of planes in phase space to allow a trajectory connecting the initial and final states to relax to the minimum free energy path. The advantages of the FFS-type methods over other transition path and rate constant calculation methods are that no assumptions are made about ``loss of memory'' during the transition, no {\em{a priori}} knowledge is required of the steady state phase space density, and the rate constant is obtained in a simple and straightforward way. We have recently become aware that the BG method bears resemblance to the RESTART method, used for simulating telecommunications networks \cite{altamirano1,altamirano2,altamirano3} (this approach was originally introduced by Bayes \cite{bayes}). The efficiency of that method has also been analysed \cite{altamirano2}. A related method, known as Weighted Ensemble Brownian Dynamics, has been applied to protein association reactions \cite{huber}. 

The key aim of a rare event simulation technique is to calculate the rate constant, or in some cases, obtain the TPE, with enhanced efficiency, compared to brute force simulations. However, quantifying the efficiency of a particular simulation method is often difficult. Our aim in this paper is to  derive simple but accurate expressions for the computational cost and statistical accuracy of the three FFS-type methods. We define  the ``efficiency'' of the methods to be the inverse of the product of the cost and the variance in the calculated rate constant; our results then allow us to analyse the efficiency of the methods in a systematic way. From a practical point of view, we expect the expressions derived here to be of use to those carrying out simulations in two ways.  Firstly, when faced with a rare event problem, one often has a limited  amount of computer time available, and specific requirements as to the desired accuracy of the calculated rate constant. Analytical expressions for the cost and statistical accuracy would  allow one to estimate, before beginning the calculation, whether the desired accuracy can be obtained within the available time, and thus to make an informed decision as to which, if any, method to use. Secondly, after completing a rate constant calculation, one needs to obtain error bars on the resulting value - this is especially important for rare events, where both experimental and simulation results can be highly inaccurate. In general, error estimation requires 
 the calculation to be repeated several times, which is computationally expensive. However, if analytical expressions were available for the statistical accuracy, in terms of quantities which were already  measured during the rate constant calculation, one could obtain the error bars on the predicted rate constant,   to within reasonable accuracy, without the need for lengthy  additional calculations. In this paper, we derive such analytical expressions. 

Approximate expressions are derived for the cost, in simulation steps, and for the variance in the calculated rate constant, for the three FFS-type methods. We initially treat the simple case where all trials fired from one interface have equal probability of succeeding. We then move on to the more realistic case where the probability of reaching the next interface depends on the identity of the starting point. To this end, we include in our calculations the ``landscape variance'' - the variance in the probability of reaching the next interface, due to the characteristic ``landscape'' for this particular rare event problem. 
 Our expressions are functions of user-defined parameters, such as the number of trial runs per point at a particular interface, as well as parameters characterizing the rare event problem itself, such as the probability that a trial run succeeds in reaching the next interface.

  We analyse the efficiency of the three methods as a function of the  parameters, for a ``generalized'' model system.  We find that the optimum efficiency is similar for all three methods, but that the effects of changing the parameter values are very different for the three methods.  In particular, the BG method performs well only within a narrow range of parameter values, while the FFS and RB methods are more robust to changes in the parameters. The RB method has consistently lower efficiency, due to its requirement for an acceptance/rejection step - however, RB may be more suitable for applications where analysis of transition paths as well as rates is needed, or where storage of configurations is very expensive. 

 To test the accuracy of our predictions in the context of  a real simulation problem, we then apply the three FFS-type methods to the  two-dimensional non-equilibrium rare event problem proposed by Maier and Stein \cite{ms_pre93,ms_jsp96,l_prl99}. We measure the computational cost of the methods and the variance in the final value of the rate constant, and we compare these to the cost and variance predicted by the expressions derived earlier. We find that the expressions give remarkably good predictions, both for the cost and the variance. This suggests that the expressions can, indeed, be used to give accurate and easy-to-calculate error estimates for real simulation problems.

In Section \ref{ffs_recap}, we briefly describe the three FFS-type methods. Expressions for the computational cost and for the statistical error in the calculated rate constant are derived in Section \ref{efficiency}. In Section \ref{sec_ms}, these expressions are shown to be accurate for the  two-dimensional non-equilibrium rare event problem proposed by Maier and Stein. Finally, we discuss our conclusions in  Section \ref{discuss}.

\section{Background: FFS-type methods}\label{ffs_recap}
\begin{figure}[h]
\begin{center}
{\rotatebox{0}{{\includegraphics[scale=0.3,clip=true]{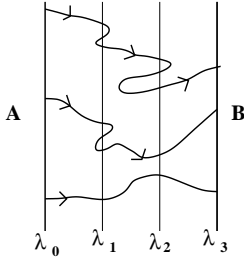}}}}
\caption{Schematic illustration of the definition of regions $A$ and $B$ and the interfaces  $\lambda_0 \dots \lambda_n$ (Here, $n=3$). Three transition paths are shown.\label{fig1} }
\end{center}
\end{figure}

The FFS-type methods use the ``effective positive flux'' expression for the rate constant, which was rigorously derived by Van Erp {\em{et al}} \cite{vanerpphd,vanerp,moroni,moroniphd,vanerp2005}. The rare event consists of a transition between two regions of phase space $A(x)$ and $B(x)$, where $x$ denotes the coordinates of the phase space. The transition occurs much faster than the average waiting time in the $A$ state. We assume that a parameter $\lambda(x)$ can be defined, such that $\lambda < \lambda_A$ in $A$ and $\lambda > \lambda_B$ in $B$. A series of values of $\lambda$, $\lambda_0 \dots \lambda_n$, are chosen such that $\lambda_0 \equiv \lambda_A$, $\lambda_n \equiv \lambda_B$ and $\lambda_i < \lambda_{i-1}$. These must constitute  a series of  non-intersecting surfaces in phase space, such that any transition path leading from $A$ to $B$ passes through each surface in turn. This is illustrated in Figure \ref{fig1}.

\noindent The rate constant $k_{AB}$ can be expressed as \cite{vanerp2005}
\begin{equation}\label{eq1}
k_{AB} = \frac{{\overline{\Phi}}_{A,n}}{{\overline{h}}_{\mathcal{A}}} = \frac{{\overline{\Phi}}_{A,0}}{{\overline{h}}_{\mathcal{A}}} P(\lambda_n|\lambda_0) .
\end{equation}
In Eq. (\ref{eq1}), $h_{\mathcal{A}}$ is a   history-dependent function describing whether the system was more recently in $A$ or $B$:  $h_{\mathcal{A}}=1$ if the system was more recently in $A$ than in $B$, and $h_{\mathcal{A}}=0$ otherwise \cite{vanerpphd,vanerp,vanerp2005}.  The over-bar denotes a time average. ${\Phi}_{A,j}$ is the flux of trajectories  with $h_{\mathcal{A}}=1$ that cross $\lambda_j$ for the first time - {\em{i.e.}} those trajectories that cross $\lambda_j$, having been in $A$ more recently than any previous crossings of $\lambda_j$.   $P(\lambda_{j}|\lambda_{i})$ is the probability that a trajectory that comes from $A$ and crosses $\lambda_{i}$ for the first time will subsequently reach $\lambda_{j}$ before returning to $A$: thus $P(\lambda_n|\lambda_0)$ is the probability that a trajectory that leaves $A$ and crosses $\lambda_0$ will subsequently reach $B$ before returning to $A$. Eq.(\ref{eq1}) states that the flux of trajectories from $A$ to $B$ can be expressed as the flux leaving $A$ and crossing $\lambda_0$, multiplied by the probability that one of these trajectories will subsequently arrive at $B$ rather that returning to $A$.   $P(\lambda_n|\lambda_0)$  can  be expressed as the product of the probabilities of reaching each successive interface from the previous one, without returning to $A$:
\begin{equation}\label{eq2}
P(\lambda_n|\lambda_0) = \prod_{i=0}^{n-1}P(\lambda_{i+1}|\lambda_{i})
\end{equation}
For simplicity of notation, in what follows, we define $P_B \equiv P(\lambda_{n}|\lambda_0)$, $p_i \equiv P(\lambda_{i+1}|\lambda_i)$, $q_i \equiv 1-p_i$  and $\Phi \equiv {\overline{\Phi}}_{A,0}/{\overline{h}}_{\mathcal{A}}$. We also use the superscript ``$e$'' to indicate an estimated value of a particular quantity.

Previously, we described in detail three different approaches - the ``forward flux sampling'' (FFS), ``branched growth'' (BG) and ``Rosenbluth'' (RB) methods -  to calculating $k_{AB}$, based on expressions (\ref{eq1}) and (\ref{eq2}) \cite{allen,long1}.  For completeness, we briefly repeat the description here.

\subsection{Forward flux sampling}
In FFS, the flux $\Phi$ is measured using a free simulation in the basin of attraction of region $A$. When the system leaves $A$ and crosses $\lambda_0$ for the first time (since leaving $A$),  its phase space coordinates are stored and the run is continued. In this way, a collection of $N_0$ points at $\lambda_0$ is generated, after which the simulation run is terminated. 

The probabilities $p_i$ are then estimated using a trial run procedure. Beginning with the collection of points at $\lambda_0$, a large number $M_0$ of trials are carried out. For each trial, a point  is selected at random from the collection at $\lambda_0$. This point is used to initiate a simulation run,  which is continued until the system either crosses the next interface $\lambda_1$, or re-enters $A$.  If $\lambda_1$ is reached, the final point of the run is stored in a new collection. After $M_0$ trials, $p_0$ is given by $N_s^{(0)}/M_0$, where $N_s^{(0)}$ is the number of trials which reached $\lambda_1$. The probability $p_1$ is then estimated in the same way: the new collection of points at $\lambda_1$ is used to initiate $M_1$ trial runs to $\lambda_2$ (or back to $A$), generating  a new collection of points at $\lambda_2$, and so on. Finally, the rate constant is obtained using Eqs (\ref{eq1}) and (\ref{eq2}).

FFS generates transition paths according to their correct weights in the TPE \cite{allen,long1}. In order to analyse these transition paths, one begins with the collection of trial runs which arrive at $\lambda_B$ from $\lambda_{n-1}$ and traces back the sequence of connected partial paths which link them to region $A$.  The resulting transition paths are branched - {\em{i.e.}} a single point at $\lambda_0$ can be the starting point of multiple transition paths. 

\subsection{The branched growth method}

In the BG method, which was inspired by techniques for polymer sampling \cite{daan,grassberger,aldous}, branched transition paths are generated one by one, rather than simultaneously, as in FFS.  The generation of each path begins with a single point at $\lambda_0$, obtained using a simulation in the basin of attraction of $A$, as in the FFS method. This point is used to initiate $k_0$ trial runs, which are continued until they either reach $\lambda_1$, or return to $A$.   Each of the  $N_s^{(0)}$ end points at $\lambda_1$ becomes a starting point for $k_1$ trial runs to $\lambda_2$ or back to $A$. Each of the $N_s^{(1)}$ successful trial runs to  $\lambda_2$ initiates $k_2$ trials to $\lambda_3$, and so on until $\lambda_n$ is reached.  An estimate $P_B^e$  of $P_B$ is obtained as  the total number of branches that eventually reach $\lambda_n$, divided by the total possible number: $P_B^e = N_s^{(n-1)}/\prod_{i=0}^{n-1} k_i$. If, at any interface, no trials were successful, $P_B^e=0$. To generate the next branching path, we obtain a new starting point at $\lambda_0$ from the simulation in the basin of attraction of $A$. After many branching paths have been generated, an average is taken over the  $P_B^e$ values of all the paths.   The flux $\Phi$ is meanwhile obtained from the simulation run in region $A$. The branched transition paths that are generated in the BG method are correctly weighted members of the TPE \cite{long1}. We note that the BG method bears resemblance to methods developed for telecommunication networks \cite{altamirano1,altamirano2,altamirano3} and to a method used for protein association \cite{huber}.

\subsection{The Rosenbluth method}\label{sec_ros}
The RB path sampling method is related to the Rosenbluth scheme for sampling polymer configurations \cite{daan,rosenbluth,fms}. The RB method generates unbranched transition paths, one at a time. An initial point at $\lambda_0$ is obtained using a simulation in the $A$ basin, which is continued until the trajectory crosses $\lambda_0$ for the first time, as in the FFS and BG methods. This point is used to initiate $k_0$ trials, which are continued until they either reach $\lambda_1$ or return to $A$. If $N_s^{(0)}>0$ of these trials reach $\lambda_1$, one successful trial is selected at random and its end point at $\lambda_1$ is used to initiate $k_1$ trials to $\lambda_2$ or back to $A$. Once again a successful trial is chosen at random and the process is repeated until either no trials are successful or $\lambda_n$ is reached.  The generation of the next path then begins with a new point at $\lambda_0$, obtained using the simulation run in the $A$ basin.

The Rosenbluth method as outlined above does not, however, generate paths according to their correct weights in the TPE: for correct sampling, paths must be re-weighted by a ``Rosenbluth factor''. The Rosenbluth factor for a partial path up to interface $i$ is given by: 
\begin{equation}\label{rfac}
W_i = \prod_{j=0}^{i-1} N_s^{(j)}
\end{equation}
Note that the re-weighting factor  $W_i$ depends on the number of successful trials obtained at all the previous interfaces, while generating the path up to $\lambda_i$. The correct re-weighting can be achieved using a Metropolis-type acceptance/rejection scheme \cite{daan}, in which a newly generated path is either accepted or rejected based on a comparison of its Rosenbluth factor with that of a previously generated path. Ensemble averages of any quantity of interest are then taken over all accepted paths. Here, the quantity which we wish to calculate is the probability $p_i$ that a trial run fired from $\lambda_i$ will reach $\lambda_{i+1}$, for each interface $i$. When we fire $k_i$ trial runs from $\lambda_i$, we obtain an estimate for $p_i$: $p_i^{e} \equiv N_s^{(i)}/k_i$. We require the correctly weighted ensemble average for $p_i^{e}$ at each interface $i$; we note, however, that the same procedure could also be used to calculate the ensemble average of any other property of the ensemble of paths from $\lambda_0$ to $\lambda_i$.

From a practical point of view, each interface has associated with it {\em{two}} values of $W_i$ and $p_i^{e}$. The first set of values:  $W_i^{({\rm{n}})}$ and $p_i^{e({\rm{n}})}$, are associated with the transition path that is currently being generated (the ``new'' path). $W_i^{({\rm{n}})}$ depends on the number of successful trials generated in creating this transition path as far as $\lambda_i$, and $p_i^{e({\rm{n}})}\equiv N_s^{(i)}/k_i$ depends on the number of successful trials fired from the point at $\lambda_i$ to $\lambda_{i+1}$. The other set of values, $W_i^{({\rm{o}})}$ and $p_i^{e({\rm{o}})}$, are the ``old'' values for this interface. These values correspond to the last ``acceptance'' event at this interface.

The recipe for obtaining $k_{AB}$ within the RB method is as follows. Transition paths are generated as described above. When the path generation procedure reaches $\lambda_i$,  we calculate the Rosenbluth factor $W_i^{({\rm{n}})}$ (using Eq.(\ref{rfac})) and we fire $k_i$ trial runs to obtain $p_i^{e({\rm{n}})}\equiv N_s^{(i)}/k_i$. We then calculate the ratio $W_i^{({\rm{n}})}/W_i^{({\rm{o}})}$ and  draw a random number  $0<s<1$. If $s < W_i^{({\rm{n}})}/W_i^{({\rm{o}})}$, an acceptance event takes place. In this case, the previous values of $W_i^{({\rm{o}})}$ and $p_i^{e({\rm{o}})}$ are replaced by the newly obtained values $W_i^{({\rm{n}})}$ and $p_i^{e({\rm{n}})}$.  If, however, $s > W_i^{({\rm{n}})}/W_i^{({\rm{o}})}$, a rejection occurs and   $W_i^{({\rm{o}})}$ and $p_i^{e({\rm{o}})}$ remain unchanged for this interface.  Regardless of the outcome of the acceptance/rejection step, the accumulator for the  probability $p_i^e$ is incremented by the current value of $p_i^{e({\rm{o}})}$ - this may be either a newly generated value (if an acceptance just occurred) or an old value that may have been already added to the accumulator several times (if several rejections have happened in a row). To proceed to the next interface, a  successful trial run is chosen out of those that have been newly generated, and its end point at $\lambda_{i+1}$ is used as the starting point for $k_{i+1}$ trial runs to $\lambda_{i+2}$. A corresponding acceptance/rejection step is then carried out at $\lambda_{i+1}$. We note that the ``old'' values $W_i^{({\rm{o}})}$ and $p_i^{e({\rm{o}})}$ for different interfaces need not correspond to the same transition path. After many complete transition paths have been generated, $k_{AB}$ is obtained using Eq.(\ref{eq1}), where an estimate of the flux $\Phi$ is calculated from the simulation run in region $A$. A ``pseudo-code'' corresponding to the above procedure is given in our previous publication \cite{long1}, together with a description of an alternative, ``Waste Recycling''  \cite{frenkel} re-weighting scheme.  In this paper, however, we shall consider only the Metropolis acceptance/rejection approach.

\section{Computational Efficiency}\label{efficiency}
In this section, we derive approximate expressions for the computational efficiency of the three methods.  Following Mooij and Frenkel \cite{mooij}, we use the following definition for the efficiency, ${\mathcal{E}}$:
\begin{equation}\label{eq_mooij}
{\mathcal{E}}=\frac{1}{{\mathcal{C}} {\mathcal{V}}}
\end{equation}
In Eq. (\ref{eq_mooij}), ${\mathcal{C}}$ represents the computational cost, which we define to be the average number of simulation steps, per initial point at $\lambda_0$. The statistical error in the estimated value $k_{AB}^e$ of the rate constant is represented by ${\mathcal{V}}$. Denoting the mean (expectation value) of variable $u$ by $E[u]$ and its variance by $V[u]$, we define ${\mathcal{V}}$ to be  the  variance $V[k_{AB}^e]$, per initial point at $\lambda_0$,  divided by the square of the expectation value $E[k_{AB}^e]$:
\begin{equation}\label{err11}
{\mathcal{V}} = \frac{N_0 V[k_{AB}^e]}{(E[k_{AB}^e])^2} = N_0 \frac{V[k_{AB}^e]}{k_{AB}^2}
\end{equation}
where $N_0$ is the number of starting points at $\lambda_0$ used in obtaining the estimate $k_{AB}^e$. The expectation value of $k_{AB}^e$ is, of course, the true rate constant: $E[k_{AB}^e]=k_{AB}$. The error bar for $k_{AB}^e$ is given by  $k_{AB}\sqrt{\mathcal{V}/N_0}$.

\subsection{Computational Cost}\label{cost_sect}
We define the computational cost ${\mathcal{C}}$ of a particular method to be the average number of simulation steps required by that method, per starting point at $\lambda_0$. In making this definition, we ignore any other contributions to the CPU time, such as memory storage. To estimate the value of ${\mathcal{C}}$, we  consider a generic system that makes a rare transition between states $A$ and $B$. A parameter $\lambda$ and interfaces $\lambda_0 \dots \lambda_n$ are chosen as in Section \ref{ffs_recap}.

There are two contributions to the cost  ${\mathcal{C}}$. The first is  the average cost $R$, in simulation steps, of generating one starting point at $\lambda_0$. This is related to the flux $\Phi$ from the $A$ region to $\lambda_0$ by 
 $R=1/(\Phi dt)$, where $dt$ is the simulation timestep.  

The second contribution to ${\mathcal{C}}$ is the cost of the trial run procedure. We first consider the cost $C_i$ of firing one trial run from interface $\lambda_i$. The run is continued until it reaches either the next interface $\lambda_{i+1}$ (with probability $p_i$), or the boundary $\lambda_A$ of region $A$ (with probability $q_i$). We make the assumption that the average length (in simulation steps) of a trajectory from interface $\lambda_i$ to another interface $\lambda_j$ is linearly proportional to $|\lambda_j-\lambda_i|$, with proportionality constant $S$. $C_i$ is then given by:
\begin{equation}\label{cost1}
C_i = S\left[p_i(\lambda_{i+1}-\lambda_i) + q_i(\lambda_{i}-\lambda_A)\right]
\end{equation}
The basis for the assumption of linearity in Eq.(\ref{cost1}) is  that we suppose that  the system undergoes one-dimensional diffusion along the $\lambda$ coordinate in the presence of a ``drift force'' of fixed magnitude.   For an equilibrium system, the origin of the drift force is the free energy barrier. Farkas and F{\"{u}}l{\"{o}}p have presented analytical solutions  \cite{farkas} for the mean time to capture for a particle undergoing one-dimensional diffusion with constant drift force, in the presence of two absorbing boundaries. In Appendix \ref{pl_app}, we show how these results lead to Eq.(\ref{cost1}). Eq.(\ref{cost1}) is shown to be valid  for the two-dimensional Maier-Stein problem in Section \ref{sec_ms_params} (Figure \ref{ms_udcosts}).

\subsubsection*{Expressions for the cost}

Given Eq.(\ref{cost1}), we can compute the average cost ${\mathcal{C}}$ per starting point at $\lambda_0$ of the three methods. 

 In FFS,  we make  $M_i$ trial runs from interface $i$ and, providing at least one of these is  successful, we proceed to the next interface $i+1$. In practice, $M_i$ is expected to be large enough that at least one trial run reaches $\lambda_{i+1}$. In this case, the expected cost per starting point at $\lambda_0$  is:
\begin{equation}\label{cost_ffs11}
{\mathcal{C}}^{\rm{ffs}}=R  +  \frac{1}{N_0}\sum_{i=0}^{n-1} M_iC_i
\end{equation}
Defining $k_i$ such that $k_i = M_i/N_0$, Eq.(\ref{cost_ffs11}) can be rewritten as:
\begin{equation}\label{cost_ffs21}
{\mathcal{C}}^{\rm{ffs}}=R   + \sum_{i=0}^{n-1} k_iC_i
\end{equation}
If, however, $M_i$ is small, we must take account of the possibility that none of the trial runs from $\lambda_i$ reach $\lambda_{i+1}$. In this case, the FFS procedure is terminated at interface $i$ and the cost is accordingly reduced. Since the probability of reaching interface $i>0$ is  $\prod_{j=0}^{i-1}\left(1-q_j^{M_j}\right)$ (this is the probability that at least one trial is successful at all interfaces $j<i$), Eq.(\ref{cost_ffs21}) is replaced by:
\begin{equation}\label{cost_ffs22}
{\mathcal{C}}^{\rm{ffs}}=R  + k_0 C_0 + \sum_{i=1}^{n-1} \left[ k_iC_i \prod_{j=0}^{i-1}\left(1-q_j^{N_0 k_j}\right)\right]
\end{equation}
Although the cost is reduced by failing to reach later interfaces, this of course results in a less accurate prediction of the rate constant, since the terminated FFS calculation makes no contribution to the estimate of $p_i$ for later interfaces. This will be reflected in our expression for the statistical error in Section \ref{err_sect}.

We now turn to the BG method. Here, we generate a ``branching tree'' of paths, with $N_i$ points at interface $i$ originating from a single point at $\lambda_0$. We fire $k_i$ trial runs for each of these $N_i$ points.  The average value of $N_i$ is:
\begin{equation}\label{nni}
\overline{N_i} = \prod_{j=0}^{i-1}p_j k_j \qquad (i>0)
\end{equation}
Of course $\overline{N_0}=1$. The average cost per starting point at $\lambda_0$ is therefore:
\begin{eqnarray}\label{cost_bg}
{\mathcal{C}}^{\rm{bg}}&=&R + \sum_{i=0}^{n-1} k_i C_i \overline{N_i}\\\nonumber &=& R +  k_0 C_0 + \sum_{i=1}^{n-1} \left[ k_i C_i \prod_{j=0}^{i-1}p_j k_j\right]
\end{eqnarray}

Finally, we come to the RB method. In this algorithm,  we generate unbranched paths by firing $k_i$ trials from interface $i$, choosing one successful trial at random and proceeding to interface $i+1$. If no trial runs are successful, we start again with a new  point at $\lambda_0$. The probability of reaching  interface $i>0$ is $\prod_{j=0}^{i-1}\left(1-q_j^{k_j}\right)$. The cost of the RB method, per starting point at $\lambda_0$, is therefore:
\begin{equation}\label{cost_ros}
{\mathcal{C}}^{\rm{rb}}=R  + k_0 C_0 + \sum_{i=1}^{n-1}\left[ k_i C_i \prod_{j=0}^{i-1}\left(1-q_j^{k_j}\right)\right]
\end{equation}
Once again, the ``price'' of failing to reach later interfaces will be paid in the form of an increased variance in the calculated rate constant. The effect of the Metropolis acceptance/rejection step in the RB method  appears only in the variance in $k_{AB}^e$ (Section \ref{err_sect}), and not in the cost.

\subsubsection*{Illustration}

\begin{figure}[h]
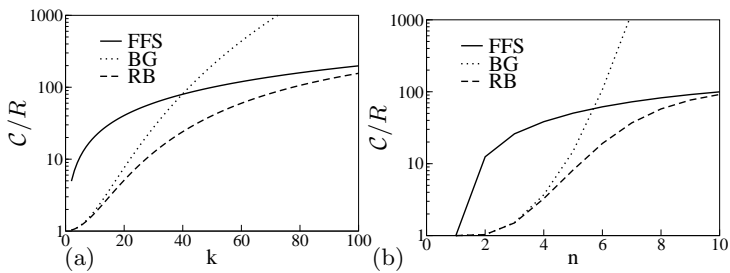

\begin{center}
\makebox[-20pt][l]{(a)}{\rotatebox{0}{{\includegraphics[scale=0.19,clip=true]{fig2a.eps}}}}\makebox[0pt][l]{(b)}{\rotatebox{0}{{\includegraphics[scale=0.19,clip=true]{fig2b.eps}}}}
\caption{ Cost ${\mathcal{C}}/R$, for evenly spaced interfaces, $p_i=p$, $k_i=k$, $R=S$, $N_0=1000$ and $P_B=10^{-8}$. (a): ${\mathcal{C}}/R$ as a function of $k$, for $n=5$. (b): ${\mathcal{C}}/R$ as a function of $n$, for $k=25$.  \label{cost_fig} }
\end{center}
\end{figure}

For the purposes of illustration, let us consider a hypothetical rare event problem for which $\lambda_0=\lambda_A=0$ and $\lambda_n=\lambda_B=1$. We suppose that the interfaces are evenly spaced in $\lambda$, have equal values of $p_i$, and that the firing parameter $k_i$  is the same at each interface: {\em{i.e.}} $\lambda_i = i/n$, $p_i = P_B^{1/n}$ (from Eq.(\ref{eq2})) and $k_i = k$. We also suppose that $R=S$ and $N_0=1000$. The resulting values of the cost ${\mathcal{C}}$, obtained from Eqs (\ref{cost_ffs22}), (\ref{cost_bg}) and (\ref{cost_ros}), are plotted in Figure \ref{cost_fig}a and b as functions of $k$ and $n$. In the regime of small $k$ or small $n$ (implying small $p$), the BG and RB methods converge, while the cost of the FFS method is higher. This is because, for BG and RB, the probability of reaching later interfaces is low and the cost is dominated by the trial runs fired from early interfaces. The FFS procedure is less likely to be terminated at early interfaces (note the factor of $1-q_i^{N_0k_i}$ in Eq.(\ref{cost_ffs22}) as opposed to $1-q_i^{k_i}$ in Eq. (\ref{cost_ros})), and is therefore more expensive, per initial point at $\lambda_0$. In the regime of large $k$  or large $n$ (implying large $p$), a different scenario emerges. Here, the BG method becomes by far the most expensive, with a cost that increases dramatically with increasing $k$ or $n$. This effect is due to the rapidly increasing number of branches per starting point at $\lambda_0$. In this regime, the FFS and RB methods converge to the same cost, since Eqs (\ref{cost_ffs22}) and (\ref{cost_ros}) become equivalent when  $1-q^k \approx 1-q^{N_0k} \approx 1$.

\subsection{Statistical Error}\label{err_sect}
We now turn to the relative variance ${\mathcal{V}}$ in the estimated value $k_{AB}^e$ of the rate constant, per starting point at $\lambda_0$. $k_{AB}^e$ is the product of the estimated flux through $\lambda_0$, multiplied by the estimated probability of subsequently reaching $B$: $k_{AB }^e = \Phi^e P_B^e$ (Eq.(\ref{eq1})). 

In this paper, we shall  ignore the error in $\Phi^e$. $\Phi^e$ is obtained  by  carrying out a simulation run in the basin of attraction of $A$ and measuring the  average number of simulation steps between successive crossings of $\lambda_0$ (coming directly from $A$). As long as $\lambda_0$ is positioned close enough to the $A$ region, the simulation run in $A$ can be made long enough to estimate $\Phi$ with high accuracy, with a computational cost that is minimal compared to the cost of estimating $P_B$. We therefore obtain:
\begin{equation}\label{err1}
{\mathcal{V}} \equiv N_0\frac{V[k_{AB}^e]}{(E[k_{AB}^e])^2} \approx N_0\frac{\Phi^2 V[P_{B}^e]}{(\Phi E[P_{B}^e])^2} =  N_0\frac{V[P_{B}^e]}{P_{B}^2} 
\end{equation}
In Eq.(\ref{err1}), we have used the general relation \cite{riley}
 \begin{equation}\label{veq3}
V[ax] = a^2V[x]
\end{equation}
where $a$ is a constant.

In what follows, we shall make the important assumption that the numbers $N_s^{(i)}$ of successful trial runs at different interfaces $i$ are {\em{uncorrelated}} - {\em{i.e.}} that if, during the generation of a transition path, one is particularly successful or unsuccessful at interface $i$, this will have no effect on the chances of success at interface $i+1$. In reality, of course, there will be correlation between interfaces, especially if the interfaces are closely spaced or the system dynamics have a large degree of ``memory''. We expect this assumption to be the major limiting factor in the applicability of our results to real systems; however, as we shall see in Section \ref{sec_ms}, the results are surprisingly accurate for the two-dimensional Maier-Stein problem. We expect that the expressions derived here could be modified to include the effects of correlations between interfaces; for highly correlated systems this may prove necessary.

\subsubsection*{Expressions for the variance}\label{pb_sec}
The basis of our analysis  is the fact that on firing $k_i$ trial runs from interface $i$, the number of successful trials $N_s^{(i)}$ is binomially distributed \cite{riley}, with mean
\begin{equation}\label{simp_e}
E[N_s^{(i)}] = k_i p_i
\end{equation}
 and variance
\begin{equation}\label{simp}
V[N_s^{(i)}] = k_i p_i q_i
\end{equation}
For now, we assume that all trial runs fired from interface $\lambda_i$ have equal probability $p_i$ of reaching $\lambda_{i+1}$. This assumption will later be relaxed. We shall need to express the variance in $P_B^e$ in terms of the variance $V[p_i^e]$ in the estimated values of $p_i$ at each interface. To do this, we recall that $P_B^e = \prod_{i=0}^{n-1}p_i^e$ (Eq.(\ref{eq2})), and we make use of the following relation \cite{riley}:
\begin{equation}\label{veq2}
V[f(x,y, \dots)] = \left(\frac{\partial f}{\partial x}\right)^2V[x] +\left(\frac{\partial f}{\partial y}\right)^2V[y] + \dots 
\end{equation}
where $f(x,y, \dots)$ is a function of multiple uncorrelated variables $x,y, \dots$ and the partial derivatives are evaluated with all variables at their mean values. By ``uncorrelated variables'' we mean that the covariance ${\mathrm{Cov}}[u,v]=0$ for all pairs of variables $u$ and $v$. Identifying $x,y \dots$ with $p_i^e, p_{i+1}^e \dots$ and taking $f(p_0^e \dots p_{n-1}^e) = \prod_{i=0}^{n-1}p_i^e$, we find that ${\partial f}/{\partial p_i^e} = [\prod_{j=0}^{n-1}p_j^e]/p_i = P_B^e/p_i^e$, so that
\begin{equation}\label{est1}
V[P_B^{e}] =  \sum_{i=0}^{n-1}E\left[\frac{P_B^e}{p_i^e}\right]^2 V[p_i^{e}] \approx P_B^2 \sum_{i=1}^{n}\frac{V[p_i^{e}]}{p_i^2}
\end{equation}

We now use the above results to calculate ${\mathcal{V}}$ for the FFS method. In this method,  we begin with a collection of $N_0$ points at $\lambda_0$. For each interface, $p_i^e$ is obtained by firing $M_i\equiv N_0k_i$ trial runs:  $p_i^e=N_s^{(i)}/M_i$, where $N_s^{(i)}$ is the number of trials which reach $\lambda_{i+1}$. Using  Eq.(\ref{veq3}), $V[p_i^{e}]=V[N_s^{(i)}]/M_i^2$. Using  Eq.(\ref{simp}), we find that $V[N_s^{(i)}] = M_i p_i q_i$. Noting also that $E[p_i^e] = p_i$ and using Eq.(\ref{est1}), we obtain 
\begin{equation}\label{var_ffs1}
V^{\rm{ffs}}[P_B^{e}] = P_B^2 \sum_{i=0}^{n-1} \frac{q_i}{p_i M_i} =\frac{P_B^2}{N_0}\sum_{i=0}^{n-1} \frac{q_i}{p_i k_i} 
\end{equation}
and from Eq.(\ref{err1})
\begin{equation}\label{ffs_imp1}
{\mathcal{V}}^{\rm{ffs}} = \sum_{i=0}^{n-1} \frac{q_i}{p_i k_i} 
\end{equation}
As for the cost calculation, we have assumed that $M_i$ is large enough that there is always at least one trial run which reaches the next interface. If this is not the case, we must also take account of the possibility that interfaces $i>0$ may not be reached. The probability of reaching interface $i>0$ is  $\prod_{j=0}^{i-1}\left(1-q_j^{M_j}\right)$, so that
\begin{eqnarray}\label{test1}
V[p_i^e] =  \frac{p_iq_i\left(1-q_i^{M_i}\right)}{M_i\prod_{j=0}^{i}\left(1-q_j^{M_j}\right)}
\end{eqnarray}
Eq.(\ref{test1}) is written in this form so that for $i=0$, we recover $V[p_0^e] =  p_iq_i/M_i$. Eqs (\ref{var_ffs1}) and (\ref{ffs_imp1}) must then be replaced by:
\begin{equation}\label{var_ffs}
V^{\rm{ffs}}[P_B^{e}] = P_B^2 \left[\sum_{i=0}^{n-1} \frac{q_i\left(1-q_i^{M_i}\right)}{p_i M_i \prod_{j=0}^{i}\left(1-q_j^{M_j}\right)}\right] 
\end{equation}
and
\begin{equation}\label{ffs_imp}
{\mathcal{V}}^{\rm{ffs}} = \sum_{i=0}^{n-1} \frac{q_i}{p_i k_i} \left[\frac{1-q_i^{N_0k_i}}{\prod_{j=0}^{i}\left(1-q_j^{N_0k_j}\right)}\right] 
\end{equation}

We now turn to the BG method. Here,  we begin with a single point at $\lambda_0$. From this point, we generate a branching ``tree'' of paths connecting $A$ to $B$. The value of $P_B$ is estimated by 
\begin{equation}\label{pb11}
P_B^e=\frac{N_s^{(n-1)}}{\prod_{i=0}^{n-1} k_i}
\end{equation}
where $N_s^{(n-1)}$ is the total number of trials reaching $\lambda_n\equiv \lambda_B$. We denote the number of points in the branching tree at interface $i$ by $N_i$. For a given number $N_{n-1}$ of points at $\lambda_{n-1}$, the total number of trials fired is $N_{n-1} k_{n-1}$ and the variance in $N_s^{(n-1)}$ is $V[N_s^{(n-1)}|N_{n-1}]=N_{n-1}k_{n-1}p_{n-1}q_{n-1}$ (using Eq.(\ref{simp})). However, the situation is complicated by the fact that $N_{n-1}$ itself varies; in fact, $N_{n-1}$ is simply the number of successful trial runs reaching $\lambda_{n-1}$ from $\lambda_{n-2}$, and in general: 
\begin{equation}\label{nneq}
N_{i}=N_s^{(i-1)} \qquad \qquad [i>0] 
\end{equation}

At this point, we need to calculate the variance in a quantity $Y$ which is conditional upon the value of another quantity $X$. Here, and several times in the rest of the paper, we will use the general relation 
\begin{equation}\label{veq1}
V[Y] = E\left[V[Y|X]\right] + V\left[E[Y|X]\right]
\end{equation}
where the mean and variance on the r.h.s. of Eq.(\ref{veq1}) are taken over the distribution of values of $X$. Since $E[N_s^{(n-1)}|N_{n-1}]=N_{n-1} k_{n-1} p_{n-1}$,
\begin{eqnarray}
V\left[E\left[N_s^{(n-1)}|N_{n-1}\right]\right]&=&k_{n-1}^2 p_{n-1}^2 V\left[N_{n-1}\right]\\\nonumber &=&k_{n-1}^2 p_{n-1}^2 V\left[N_s^{(n-2)}\right] 
\end{eqnarray}
(using Eqs (\ref{veq3}) and (\ref{nneq})). We also know that 
\begin{eqnarray}
E\left[V\left[N_s^{(n-1)}|N_{n-1}\right]\right]&=& k_{n-1}p_{n-1}q_{n-1}E\left[N_{n-1}\right]\\\nonumber &=& k_{n-1}p_{n-1}q_{n-1}\prod_{i=0}^{n-2}k_i p_i
\end{eqnarray}
so that
\begin{eqnarray}\label{bgvar21}
V[N_s^{(n-1)}] =q_{n-1}\prod_{i=0}^{n-1}k_i p_i + k_{n-1}^2 p_{n-1}^2 V\left[N_s^{(n-2)}\right] 
\end{eqnarray}
Using the same arguments, we can generalize Eq.(\ref{bgvar21}) to 
\begin{eqnarray}\label{bgvar2}
V[N_s^{(i)}]  &= q_i\prod_{j=0}^{i}k_jp_j + k_i^2 p_i^2 V[N_s^{(i-1)}]&\, [i>0]\\\nonumber & q_ik_ip_i &\, [i=0]
\end{eqnarray}
Using Eq.(\ref{bgvar2}), we can solve Eq.(\ref{bgvar21}) recursively, to obtain $V[N_s^{(n-1)}]$. Using Eqs.(\ref{pb11}) and (\ref{veq3}), we then arrive at the variance in the estimated value of $P_B$: 
\begin{equation}\label{var_bg}
V^{\rm{bg}}[P_B^{e}] = \frac{P_B^2}{N_0} \sum_{i=0}^{n-1} \frac{q_i}{\prod_{j=0}^{i}p_j k_j}
\end{equation}
where we have divided by $N_0$ to account for the fact that $P_B^e$ is calculated by averaging results over $N_0$ starting points at $\lambda_0$. We then obtain from Eq.(\ref{err1}):
\begin{equation}\label{bg_imp}
{\mathcal{V}}^{\rm{bg}} =  \sum_{i=0}^{n-1} \frac{q_i}{\prod_{j=0}^{i}p_j k_j}
\end{equation}

Finally, let us derive the equivalent expression for the RB method. Here, we again use Eq.(\ref{est1}). If we ignore for the moment the effect of the acceptance rejection step, we can use Eqs.(\ref{simp}) and (\ref{veq3}) to obtain an expression for the variance in $p_i^{e}$:
\begin{eqnarray}\label{ros1111}
V^{\rm{rb}}[p_i^{e}] =\frac{p_iq_i}{N_0 k_i}\frac{\left(1-q_i^{k_i}\right)}{\prod_{j=0}^{i}\left(1-q_j^{k_j}\right)}
\end{eqnarray}
where we have taken account of the fact that the probability of reaching interface $i>0$ is $\prod_{j=0}^{i-1}\left(1-q_j^{k_j}\right)$, and that the $p_i^e$ value is averaged over $N_0$ separate path generations. Eq.(\ref{ros1111}) is very similar to the FFS result, Eq.(\ref{test1}). 

The Metropolis acceptance/rejection step (described in Section \ref{ffs_recap}) increases the variance in $p_i^{e}$.  On reaching interface $i$,  we fire $k_i$ trials and obtain an estimate $p_i^{e,({\rm{n}})}=N_s^{(i)}/k_i$. We either accept or reject this estimate. If we reject, $p_i^{e,({\rm{n}})}$ makes no contribution to the  average value of $p_i^e$ - instead, the previously accepted estimate, $p_i^{e,({\rm{o}})}$, is added to the average, even though $p_i^{e,{\rm{o}}}$  was already added to the average in the previous acceptance/rejection step. If, instead, we accept $p_i^{e,({\rm{n}})}$, it makes a contribution to $p_i^e$, and, if the subsequent estimates happen to be rejected, it may repeat this contribution multiple times. The final estimate, $p_i^{e}$, is therefore an average over all the values of $N_s^{(i)}/k_i$ that were generated, weighted by the number of times $Q$ that each of these values contributed to $p_i^{e}$:
\begin{equation}
p_i^e = \frac{\sum_{l=1}^{N_g}Q_l \left[N_s^{(i)}/k_i\right]_l}{N_g^{(i)}}
\end{equation}
where the sum is over all generated $N_s^{(i)}/k_i$ values and $N_g^{(i)}$ is the total number of these. In fact, 
\begin{equation}\label{ngeq}
N_g^{(i)} = N_0 \frac{\prod_{j=0}^{i}\left(1-q_j^{k_j}\right)}{\left(1-q_i^{k_i}\right)}
\end{equation}
since the number of times we fire trials from $\lambda_i$ is simply the number of times we begin a path generation from $\lambda_0$ and succeed in reaching $\lambda_i$.  Using Eq.(\ref{veq3}), the variance $p_i^e$ is then
\begin{equation}\label{test3}
V[p_i^e] = \frac{\sum_{l=1}^{N_g^{(i)}}Q_l^2 V\left[N_s^{(i)}/k_i\right]_l}{(N_g^{(i)})^2} = \frac{V\left[N_s^{(i)}\right]}{k_i^2(N_g^{(i)})^2}\sum_{l=1}^{N_g^{(i)}}Q_l^2
\end{equation}
(assuming that the distributions of the stochastic variables $Q_l$ and $[N_s^{(i)}]_l$ are uncorrelated). Eq.(\ref{test3}) is equivalent to:
\begin{equation}\label{test2}
V[p_i^e] =  \frac{V\left[N_s^{(i)}\right]}{k_i^2N_g^{(i)}} \sum_{Q=0}^{\infty}Q^2P(Q) = \frac{p_i q_i}{k_iN_g^{(i)}}\sum_{Q=0}^{\infty}Q^2 P(Q)
\end{equation}
In order to find the distribution $P(Q)$, we define a new variable $\theta_i$. $\theta_i$ is the probability that we accept a newly generated estimate $p_i^{e,({\rm{n}})}=N_s^{(i)}/k_i$.  $P(Q)$ is then:
\begin{eqnarray}\label{qdist}
P(Q) =& (1-\theta_i)\qquad & Q=0\\\nonumber
P(Q) = &\theta_i^2 (1-\theta_i)^{Q-1}\qquad & Q>0 
\end{eqnarray}
Eq.(\ref{qdist}) can be understood as follows: $Q=0$ corresponds to a $p_i^{e,({\rm{n}})}$ value that is generated but is immediately rejected and therefore contributes zero times to the average. This occurs with probability $1-\theta_i$.  $Q>0$ corresponds to a $p_i^{e,({\rm{n}})}$ value that is generated and accepted (with probability $\theta_i$) - the next $Q-1$ values that are generated are rejected  (with probability $(1-\theta_i)^{Q-1}$),  then finally a new value is generated which is accepted  (with probability $\theta_i$), so that the original value ceases to contribute to the average. The distribution (\ref{qdist}) has the property that \cite{gradshteyn} 
\begin{equation}
\sum_{Q=0}^\infty Q^2 P(Q) = \frac{2-\theta_i}{\theta_i}
\end{equation}
so that Eq.(\ref{test2}) for the variance in $p_i^e$ per point at $\lambda_i$ becomes
\begin{equation}
V[p_i^e] = \frac{p_i q_i}{k_iN_g^{(i)}}\left[\frac{2-\theta_i}{\theta_i}\right]
\end{equation}
Using Eq.(\ref{ngeq}), we obtain:
\begin{eqnarray}\label{ros111}
V^{\rm{rb}}[p_i^{e}] =\frac{p_iq_i}{N_0 k_i}\left[\frac{(2-\theta_i)}{\theta_i}\right]\frac{\left(1-q_i^{k_i}\right)}{\prod_{j=0}^{i}\left(1-q_j^{k_j}\right)}
\end{eqnarray}
Comparing to Eq.(\ref{ros1111}), we see that the effect of the acceptance/rejection step is to multiply $V[p_i^{e}]$ by a factor $(2-\theta_i)/\theta_i$. Using Eq.(\ref{est1}), the  relative variance in $P_B^e$ is:
\begin{equation}\label{var_ros}
\frac{V^{\rm{rb}}[P_B^{e}]}{ P_B^2} = \frac{1}{N_0}\sum_{i=0}^{n-1} \frac{q_i}{p_ik_i}\frac{(2-\theta_i)}{\theta_i}\frac{\left(1-q_i^{k_i}\right)}{\prod_{j=0}^{i}\left(1-q_j^{k_j}\right)}
\end{equation}
so that using Eq.(\ref{err1}),
\begin{equation}\label{rb_imp}
{\mathcal{V}}^{\rm{rb}} =  \sum_{i=0}^{n-1}\frac{q_i}{p_i k_i}\frac{(2-\theta_i)}{\theta_i}\frac{\left(1-q_i^{k_i}\right)}{\prod_{j=0}^{i}\left(1-q_j^{k_j}\right)}
\end{equation}
We show in Appendix \ref{app_theta} that the acceptance probability $\theta_i$  for $i>0$ [note that $\theta_0=1$] can be approximated as:
\begin{equation}\label{theta_1}
\theta_i  = \frac{1}{2}-\frac{\sqrt{\pi}}{4}\left[2{\rm{Erf}}\left(\frac{\sigma_i}{2}\right)-1\right] \qquad (i>0)
\end{equation}
where ${\rm{Erf}}(x)$ is the error function: ${\rm{Erf}}(x)=(2/\sqrt{\pi})\int_0^x e^{-t^2}dt$, and $\sigma_i$ is given by:
\begin{equation}\label{theta_2}
\sigma_i^2 = \sum_{j=0}^{i-1} \left[\frac{(1-q_j^{k_j})q_j}{k_jp_j}-q_j^{k_j}\right] 
\end{equation}
Eqs (\ref{theta_1}) and (\ref{theta_2}) can be substituted into Eq.(\ref{rb_imp}) to give a complete expression for the relative variance in the estimated rate constant for the RB method.

\subsubsection*{Illustration}

\begin{figure}[h]
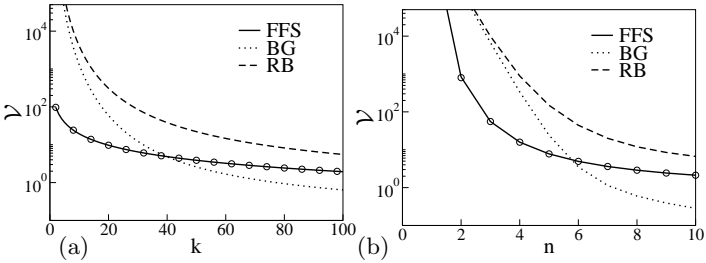

\begin{center}
\makebox[-20pt][l]{(a)}{\rotatebox{0}{{\includegraphics[scale=0.19,clip=true]{fig3a.eps}}}}\makebox[0pt][l]{(b)}{\rotatebox{0}{{\includegraphics[scale=0.19,clip=true]{fig3b.eps}}}}
\caption{Relative variance ${\mathcal{V}}$, for $p_i=p$, $k_i=k$ and $P_B=10^{-8}$. The circles show the function $\sum_{i=0}^{n-1} q_i/(p_i k_i)$. (a): ${\mathcal{V}}$ as a function of $k$, for $n=5$. (b): ${\mathcal{V}}$ as a function of $n$, for $k=25$. \label{var_fig} }
\end{center}
\end{figure}

Returning to the hypothetical rare event problem with evenly spaced interfaces introduced above, Figure  \ref{var_fig} shows ${\mathcal{V}}$ as a function of $k$ (for $n=5$) and of $n$ (for $k=25$), for $p_i=p=P_B^{1/n}$, $k_i=k$, $N_0=1000$ and $P_B=10^{-8}$. The circles show the limiting form $\sum_{i=0}^{n-1} q_i/(p_i k_i)$, which is in good agreement with the FFS results, since $1-q^{N_0k}\approx 1$. For small $k$ or small $n$ (small $p$), the RB and BG results tend to converge, since the probability of reaching later interfaces is small and the results are dominated by the early interfaces. In this regime,  the FFS method gives the smallest variance, since the chance of terminating the trial run procedure at early interfaces is lower than for the other methods.

It is interesting to compare  expressions (\ref{ffs_imp}), (\ref{bg_imp}) and (\ref{rb_imp}).  All three expressions are of the form
\begin{equation}
{\mathcal{V}} = \sum_{i=0}^{n-1} \frac{q_i}{p_i k_i X_i}
\end{equation}
However, $X_i$ takes different forms for the three methods: 
\begin{equation}
X_i^{\rm{ffs}} = \frac{\prod_{j=0}^{i} (1-q_j^{N_0 k_j})}{(1-q_i^{N_0 k_i})}
\end{equation}
\begin{equation}
X_i^{\rm{bg}} = \prod_{j=0}^{i} p_j k_j
\end{equation}
and
\begin{equation}
X_i^{\rm{rb}} = \frac{\theta_i}{(2-\theta_i)}\frac{\prod_{j=0}^{i-1} (1-q_j^{k_j})}{(1-q_i^{k_i})}
\end{equation}
We note that $X_i^{\rm{ffs}} > X_i^{\rm{rb}}$, so that ${\mathcal{V}}^{\rm{ffs}}$ is always less than ${\mathcal{V}}^{\rm{rb}}$, even for $\theta_i=1$. Both $X_i^{\rm{ffs}}$ and $X_i^{\rm{rb}}$ are always less than unity: ${\mathcal{V}}^{\rm{ffs}}$ approaches the limiting form  $\sum_{i=0}^{n-1} q_i/(p_i k_i)$ from above as $k_i$ increases (in fact in Fig. \ref{var_fig}a it takes this form for all $k$) and ${\mathcal{V}}^{\rm{rb}}$ approaches $\sum_{i=0}^{n-1} (2-\theta_i)q_i/(p_i k_i\theta_i)$. For the BG method, however, $X_i^{\rm{bg}}$ can increase indefinitely as $k_i$ increases, so that this method produces the smallest variance for large $k_i$, as in Figure \ref{var_fig}a. However, comparing with Figure \ref{cost_fig}, we see that this is also the regime in which the BG method becomes very expensive.

\subsubsection*{Landscape Variance}\label{int_var_sec}
So far in our analysis, we have assumed that all the points at interface $\lambda_i$ have to same $p_i$ value - {\em{i.e.}} that on firing a trial run to $\lambda_{i+1}$ we have the same probability of success, no matter which point at $\lambda_i$ we start from. In reality, this is not the case; we expect there to be a distribution of $p_i$ values among the points at each interface $\lambda_i$. We call the variance of this distribution the ``landscape variance'' $U_i$ at interface $i$, and we expect it to make a contribution to the variance in $P_B^e$. We now extend our analysis to include the potentially important effect of the landscape variance.

Let us suppose that each  point $j$ at $\lambda_i$ has an associated probability $p_i^{(j)}$ that a trial run fired from that point will reach $\lambda_{i+1}$. The distribution of $p_i^{(j)}$ values encountered during the rate constant calculation has mean $E[p_i^{(j)}]=p_i$ and variance $V[p_i^{(j)}] \equiv U_i$. Of course, the values of $U_i$ depend on the number and placement of the interfaces.

In Appendix \ref{int_var_inc}, we re-derive expressions for the relative variance in the estimated rate constant, taking into account the landscape variance. The final results are:
\begin{eqnarray}\label{ffs_imp_int}
\nonumber {\mathcal{V}}^{\rm{ffs}} = \sum_{i=0}^{n-1} && \Bigg\{\left[\frac{q_i}{p_i k_i} + \frac{U_i N_0}{p_i^2 N_i}\left(1-\frac{1}{N_0k_i}\right)\right]\\ && \times\frac{\left(1-q_i^{N_0k_i}\right)}{ \prod_{j=0}^{i}\left(1-q_j^{N_0k_j}\right)}\Bigg\}
\end{eqnarray}
where $N_i=N_0 k_{i-1}p_{i-1}$ for $i>0$ and $N_i=N_0$ for $i=0$.
\begin{equation}\label{bg_imp_int}
{\mathcal{V}}^{\rm{bg}} =  \sum_{i=0}^{n-1} \left[\frac{k_iq_ip_i + U_i \left(k_i^2-k_i\right)}{k_ip_i\prod_{j=0}^{i}p_j k_j}\right] 
\end{equation}
and
\begin{eqnarray}\label{rb_imp_int}
{\mathcal{V}}^{\rm{rb}} = \sum_{i=0}^{n-1} && \Bigg\{\left[\frac{q_i}{p_ik_i} + \frac{U_i}{p_i^2}\left(1-\frac{1}{k_i}\right)\right]\\\nonumber && \times \left[\frac{(2-\theta_i)}{\theta_i}\right]\frac{\left(1-q_i^{k_i}\right)}{ \prod_{j=0}^{i}\left(1-q_j^{k_j}\right)}\Bigg\}
\end{eqnarray}

Comparing Eqs (\ref{ffs_imp_int}), (\ref{bg_imp_int}) and (\ref{rb_imp_int}) to their equivalent forms without landscape variance, (\ref{ffs_imp}), (\ref{bg_imp}) and (\ref{rb_imp}), we see that for each interface the ``binomial'' terms of the form $p_iq_i/k_i$ are now supplemented by additional terms describing the landscape variance. In the limit of very large $k_i$, the relative variance no longer tends to zero. Instead, as $k_i \to \infty$ (for all $i$), the FFS and BG expressions (\ref{ffs_imp_int}) and (\ref{bg_imp_int}) tend to the constant value $U_0/p_0^2$,  while the RB expression (\ref{rb_imp_int}) tends to $\sum_{i=0}^{n-1} U_i/p_i^2$. While the ``binomial'' contribution  to the variance can be reduced by firing many trial runs per point, the ``landscape'' contribution can only be reduced by sampling many points. In the FFS and BG methods, branching paths are generated. For very large $k_i$, each point at $\lambda_0$ generates many points at subsequent interfaces, so that only $U_0$ remains in Eqs (\ref{ffs_imp_int}) and (\ref{bg_imp_int}) as $k_i \to \infty$. In the RB method, however, paths are not branched, so that each point at $\lambda_0$ corresponds to one (or less than one) point at each subsequent interface. In this case, as $k_i \to \infty$, all the $U_i$ values continue to contribute to ${\mathcal{V}}$.

\begin{figure}[h]
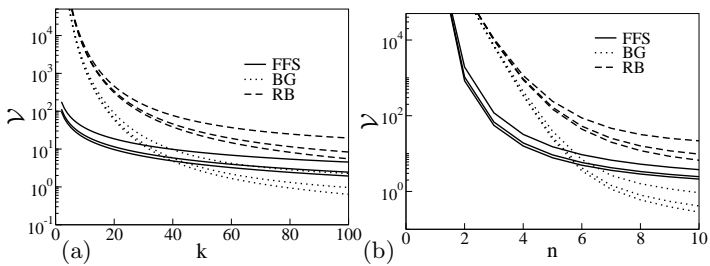

\begin{center}
\makebox[-20pt][l]{(a)}{\rotatebox{0}{{\includegraphics[scale=0.19,clip=true]{fig4a.eps}}}}\makebox[0pt][l]{(b)}{\rotatebox{0}{{\includegraphics[scale=0.19,clip=true]{fig4b.eps}}}}
\caption{Relative variance ${\mathcal{V}}$ in $k_{AB}^e$, as predicted by Eqs (\ref{ffs_imp_int}), (\ref{bg_imp_int}) and (\ref{rb_imp_int}), for the  model problem of Figs \ref{cost_fig} and \ref{var_fig}, with $P_B=10^{-8}$ and $U_i=U$. The upper curves in each group correspond to $U=5p^2/n$, the middle curves to $U=p^2/n$ and the lower curves to $U=0$. (a): ${\mathcal{V}}$ as a function of $k$, keeping $n=5$. (b): ${\mathcal{V}}$ as a function of $n$, keeping $k=25$.\label{var_fig_int} }
\end{center}
\end{figure}

In Figure \ref{var_fig_int}, we revisit the simple model problem of  Figs \ref{cost_fig} and \ref{var_fig}, adding in the effects of landscape variance. We take $U_i$ to be the same for all interfaces: $U_i=U$. We choose, somewhat arbitrarily, $U=p^2/n$ or $U=5p^2/n$. These turn out to be quite realistic values for the Maier-Stein system discussed in Section \ref{sec_ms}. Figure \ref{var_fig_int} shows the relative variance ${\mathcal{V}}$ (as in Figure \ref{var_fig}), calculated with $U=5p^2/n$ (upper curves), $U=p^2/n$ (middle curves) and $U=0$ (lower curves). Although the landscape variance does not change the general trend that ${\mathcal{V}}$ decreases as $k$ or $N$ increases, it does have the qualitative effect that ${\mathcal{V}}$ no longer tends to zero (as discussed above). Depending on the value of $U$, the quantitative effects of the landscape contribution can be very significant, especially as $k$ or $N$ becomes large.

\subsection{Efficiency}\label{eff_sect2}
\begin{figure}[h]
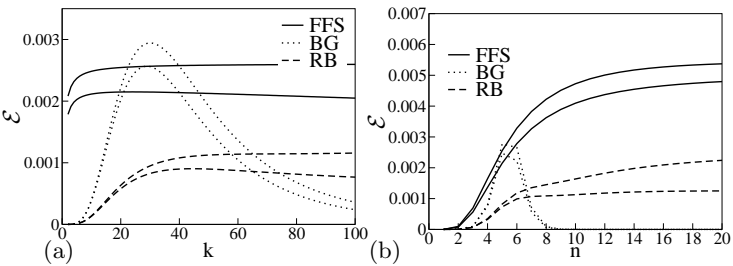

\begin{center}
\makebox[-15pt][l]{(a)}{\rotatebox{0}{{\includegraphics[scale=0.19,clip=true]{fig5a.eps}}}}\makebox[0pt][l]{(b)}{\rotatebox{0}{{\includegraphics[scale=0.19,clip=true]{fig5b.eps}}}}
\caption{Efficiency ${\mathcal{E}}$, calculated using Eq.(\ref{eq_mooij}), for the simple model of Figs \ref{cost_fig}, \ref{var_fig} and \ref{var_fig_int}. For each method, results are plotted with $U=p^2/n$ (lower curves) and $U=0$ (upper curves). (a): ${\mathcal{E}}$ as a function of $k$ for $n=5$. (b): ${\mathcal{E}}$ as a function of $n$ for $k=25$. \label{eff_fig} }
\end{center}
\end{figure}

Having calculated the computational cost and the statistical accuracy of the three  methods, we are now in a position to assess their overall computational efficiency, as defined by Eq.(\ref{eq_mooij}). Figure \ref{eff_fig} shows the efficiency of the three methods as a function of $k$ (Fig. \ref{eff_fig}a) and of $n$ (Fig. \ref{eff_fig}b), for the simple model case of Figs \ref{cost_fig}, \ref{var_fig} and \ref{var_fig_int}. Note the altered scale on the $n$ axis in comparison to Figures \ref{cost_fig} and \ref{var_fig}. For each method, the upper curve shows the results without the landscape contribution to the variance ($U=0$) and the lower curve includes a landscape contribution of $U=p^2/n$. 

Firstly, we note that  the optimum values of ${\mathcal{E}}$ are of the same order of magnitude for all three methods, although ${\mathcal{E}}$ is consistently lower for RB, due to the acceptance/rejection step. However, the dependence of the efficiency on the parameter values $k$ and $n$ is very different for the three methods. For the BG method, the efficiency shows a pronounced peak, both as a function of $k$ and of $n$. Although for an optimum choice of parameters, this method can be the most efficient,  its performance  is highly sensitive to the choice of parameters, decreasing sharply for non-optimal values of $k$ or $n$. The FFS and RB methods are much less parameter-sensitive - in fact, as long as $k$ or $n$ is not too small, the choice of parameters appears not to be at all critical for these methods. In general, Fig.\ref{eff_fig} seems to indicate that the method of choice is FFS, since this method is highly robust to changes in the parameters, is  the most efficient method at small $k$ or $n$, and remains efficient as $k$ and $n$ become large. However, this interpretation must be treated with care, since several important factors are not included in the analysis leading to Fig.\ref{eff_fig}. Firstly, our analysis does not include the effects of correlations between interfaces. This has the effect that neither the FFS or RB methods shows a maximum in efficiency as a function of $n$ in Fig.\ref{eff_fig}b. In our simple model, one can always gain more information by sampling at more closely spaced interfaces - however, in reality, correlations between interfaces are likely to make very closely spaced interfaces computationally inefficient. Another important factor to be considered is the fact that both the FFS and BG methods generate branched transition paths. In FFS, in fact, an effect analogous to ``genetic drift'' means that if the number of points in the collections at the interfaces is small enough to be of the order of the number of interfaces, then all the paths that finally reach $B$ can be expected to originate from a small number of initial points at $\lambda_0$. If there is ``memory loss'' - {\em{i.e.}} no correlations between interfaces, this may be unimportant. However, if the history of the paths is important, then  the RB method may be the method of choice, since this generates independent, unbranched paths.  Furthermore, the RB method requires much less storage of system configurations than FFS (for which a whole collection of points must be stored in memory at each interface) - for some systems, this may be a significant factor in the computational cost. 

Figure \ref{eff_fig} also shows the effects of landscape variance on the efficiency of the three methods. Including landscape variance always decreases the efficiency, but produces rather few qualitative effects for this simple model problem. It is  interesting to note, however, that in Figure \ref{eff_fig}a both the FFS and RB methods show a maximum in efficiency as a function of $k$ only when the landscape contribution is included. When  the landscape contribution is not considered, the equations predict that arbitrarily high accuracy can be obtained by firing an infinitely large number of trials from a single point. In this example, we took the landscape variance to be the same for all interfaces: $U_i=U$. However, one can easily imagine that for some systems, there is much greater variation among transition paths when they are close to the $A$ basin, while for others, paths tend to diverge as they approach $B$. In the former case, we can expect the RB and BG methods to have an advantage relative to FFS, because in these methods, relatively more points are sampled at early interfaces (since the probability of failing to complete a transition path is higher). Conversely, if the landscape variance is very large close to the $B$ basin, the BG method may be advantageous, since it samples many points at later interfaces due to its branching tree of paths.

\section{The Maier-Stein system}\label{sec_ms}
In this section, we test the expressions derived in Section \ref{efficiency} for a real rare event simulation problem. As our test case, we simulate the  two-dimensional non-equilibrium rare event problem proposed by Maier and Stein \cite{ms_pre93,ms_jsp96,l_prl99}.  This system has been extensively studied both theoretically and experimentally \cite{ms_pre93,ms_jsp96,l_prl99,ms_nat97,ms_prl97} and  was also used by  Crooks and Chandler \cite{crooks} as a test case for their non-equilibrium rare event method. We hope that the conclusions obtained for this system will also prove to be applicable to more computationally intensive rare event problems.

\begin{figure}[h]
\begin{center}
{\rotatebox{0}{{\includegraphics[scale=0.25,clip=true]{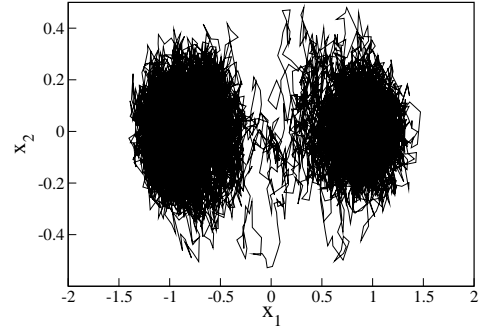}}}}
\caption{Typical trajectory for a brute-force simulation of the Maier-Stein system, with $\alpha=6.67$, $\mu=2$ and $\epsilon=0.1$.\label{ms_bf} }
\end{center}
\end{figure}

The Maier-Stein system consists of  a single particle moving with over-damped Langevin dynamics  in a two-dimensional force field.  The position vector ${(x_1,x_2)}$ of the particle satisfies the stochastic differential equation:
\begin{equation}\label{bd}
\dot{x_i} = f_i(x) + \xi_i(t)
\end{equation}
where the force field ${\bf{f}}=(f_1,f_2)$ is given by:
\begin{equation}
{\bf{f}} = \left(x_1-x_1^3-\alpha x_1x_2^2\,,\,-\mu x_2(1+x_1^2)\right)
\end{equation}
and the stochastic force ${\bf{\xi}}=(\xi_1,\xi_2)$ satisfies:
\begin{equation}
\langle \xi_i(t) \rangle = 0 \,\, ; \,\, \langle \xi_i(t+\tau) \xi_j(t)\rangle = \epsilon \, \delta(t-\tau) \delta_{ij}
\end{equation}
This system is bistable, with stable points at $(\pm 1,0)$ and a saddle point at $(0,0)$. If $\alpha \ne \mu$, the force field ${\bf{f}}$ cannot be expressed as the gradient of a potential. In this case, the system is intrinsically out of equilibrium and does not satisfy detailed balance. The parameter $\epsilon$ controls the magnitude of the stochastic force acting on the particle. For $\epsilon > 0$, the system makes stochastic transitions between the two stable states, at a rate which decreases as $\epsilon$ decreases.  Figure \ref{ms_bf} shows a typical trajectory generated by a brute-force simulation. Here, and in the rest of this Section, we use $\alpha=6.67$, $\mu=2.0$ (following Crooks and Chandler \cite{crooks}) and $\epsilon=0.1$. Eq.(\ref{bd}) is integrated numerically with timestep $\delta t=0.02$ \cite{allentildesley}. For our calculations using the FFS-type methods, we define  $\lambda ({\bf{x}}) = x_1$, $\lambda_A\equiv \lambda_0=-0.7$ and $\lambda_B\equiv \lambda_n=0.7$.

\subsection{Measuring the parameters}\label{sec_ms_params}
\begin{figure}[h]
\begin{center}
{\rotatebox{0}{{\includegraphics[scale=0.25,clip=true]{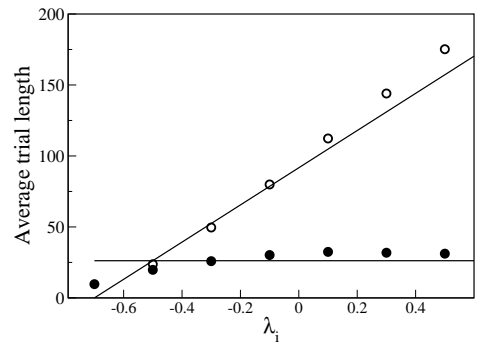}}}}
\caption{ Costs of trial runs between interfaces, for the Maier-Stein system. The average length, in simulation steps, of ``successful'' trials (to $\lambda_{i+1}$) are shown as filled circles. For these trials, $\lambda_j = \lambda_i+0.2$ and $|\lambda_i-\lambda_j|=0.2$. The average length of ``unsuccessful'' trials (to $\lambda_A=-0.7$) are shown as open circles. For these trials, $|\lambda_i-\lambda_j|=\lambda_i+0.7$. The solid lines show the linear approximation, Eq.(\ref{cost1}), with $S=131$. \label{ms_udcosts}}
\end{center}
\end{figure}

In order to test the expressions of Section \ref{efficiency}, we must measure the cost parameters $R$ and $S$, the probability $P_B$ of reaching $B$ and, for a given set of $n$ interfaces, the probabilities $\{p_i\}$ and the landscape variance values $\{U_i\}$. For most of our calculations, we used $n=7$, and the interfaces were positioned as listed in Table \ref{ms_params}. For the results of Figs \ref{cost_comp}b, \ref{var_comp}b and \ref{eff_comp}b, where $n$ was varied, we kept the interfaces evenly spaced between $\lambda_0=-0.7$ and $\lambda_n=0.7$. $R$, the cost of generating an initial point at $\lambda_0$, was measured using a simulation in region $A$ to be $R=590\pm 50$ steps. In these calculations, points at $\lambda_0$ were collected upon every $10$th crossing of $\lambda_0$ from $A$.  To measure $S$ (the proportionality constant in Eq.(\ref{cost1})), we carried out an FFS run, measuring the average length (in simulation steps) of successful and unsuccessful trials from each interface. The results are shown in Figure  \ref{ms_udcosts}. Here, the filled circles show the average length, in simulation steps, of successful trials from interface $\lambda_i$ (plotted on the x axis) to $\lambda_{i+1}=\lambda_i+0.2$. Since $|\lambda_i-\lambda_j|=0.2$ for all these trials, Eq.(\ref{cost1}) predicts that all the filled circles should have show the same average trial length. The open circles show the average length of unsuccessful trials, which begin at $\lambda_i$ and end at $\lambda_A=-0.7$, so that $|\lambda_i-\lambda_j|=\lambda_i+0.7$: Eq.(\ref{cost1}) predicts that all the open circles should lie on a straight line. Combining all the data, we obtain an average value of $S=131$ steps. This value is used to plot the solid lines in Figure \ref{ms_udcosts}. The very good agreement that is observed between the solid lines and the circles implies that  the drift-diffusion approximation, Eq.(\ref{cost1}), is reasonable for this problem. The most significant deviation occurs for the successful trial runs between $\lambda=-0.7$ and $\lambda=-0.5$; these are unexpectedly short, perhaps because the ``drift force'' is weaker in this region.

\begin{table}[h]
\begin{center}
\begin{tabular}{cccc}
Interface& $\lambda_i$ &$p_i$&$U_i$ \\
0 & -0.7 \,\,\, & $0.1144\pm 0.0001$\,\,\, & $0.00350 \pm 0.00003$\\
1 & -0.5 \,\,\, & $0.2651\pm 0.0002$\,\,\,  & $0.00368 \pm 0.00008$\\
2 &-0.3 \,\,\, & $0.3834\pm 0.0002$\,\,\,  & $0.0031 \pm 0.0003$\\
3 &-0.1 \,\,\, & $0.5633\pm 0.0003$\,\,\,  & $0.0021 \pm 0.0002$\\
4 &0.1 \,\,\, & $0.7702\pm 0.0003$\,\,\,  & $0.0008 \pm 0.0001$\\
5 &0.3 \,\,\, & $0.9152\pm 0.0002$\,\,\,  &  $0.0003 \pm 0.0001$\\
6 &0.5 \,\,\, & $0.9747\pm 0.0002$\,\,\,  & $0.00005 \pm 0.00002$\\
\end{tabular}
\end{center}
\caption{Positions of the interfaces and measured values of $\{p_i\}$ and $\{U_i\}$ for the Maier-Stein problem.\label{ms_params}}
\end{table}

Using FFS, we obtained $P_B=[4.501 \pm 0.007] \times 10^{-3}$. The values of $\{p_i\}$ were also measured (using FFS) and are given in Table \ref{ms_params}. The landscape variance $\{U_i\}$  was measured using the procedure described in Appendix \ref{int_var_meas}: after generating a correctly weighted collection of points at interface $\lambda_i$ (for example using FFS), one fires $k_i$ trials from each point $j$ and records the number of successes, $N_s^{(i)}|j$. One then calculates the variance among points $V[N_s^{(i)}]$. The intrinsic variance is given by 
\begin{equation}
U_i = \frac{V[N_s^{(i)}]/k_i - p_iq_i}{k_i-1}
\end{equation} 
Table \ref{ms_params} shows that for this problem $U_i/p_i^2$ is rather small (a maximum of 0.27 for interface 0), indicating that the landscape variance is unlikely to have important effects in this case. However, this may not be the case for more complex systems in higher dimensions.

\subsection{Testing the expressions}

We now measure directly the  cost, in simulation steps, the error in the calculated rate constant, and thus the efficiency of the three methods, for the Maier-Stein problem, and compare our simulation results to the predictions of Section \ref{efficiency}. For each method, simulations were  carried out in  a series of blocks. For FFS, a block consists of a complete FFS calculation with $N_0$ starting points. For the RB and BG methods, a block consists of $N_0$ starting points at $\lambda_0$. Each block produces a result $P_B^e$ for the probability of reaching $B$. To find $V[P_B^e]$, we calculate the variance between blocks:
\begin{equation}
V[P_B^e] = \overline{(P_B^e)^2}-(\overline{P_B^e})^2
\end{equation}
where the over-line denotes an average over the blocks.  The cost ${\mathcal{C}}$ per starting point  at $\lambda_0$ is the average number of simulation steps per block, divided by $N_0$.

\begin{figure}[h]
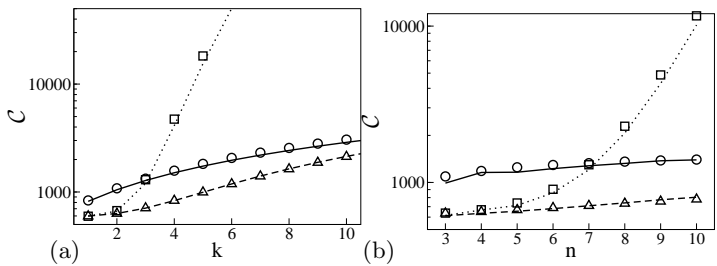

\begin{center}
\makebox[-15pt][l]{(a)}{\rotatebox{0}{{\includegraphics[scale=0.19,clip=true]{fig8a.eps}}}}\makebox[0pt][l]{(b)}{\rotatebox{0}{{\includegraphics[scale=0.19,clip=true]{fig8b.eps}}}}
\caption{Predicted and measured values of ${\mathcal{C}}$, for the Maier-Stein problem as described in Section \ref{sec_ms}. The lines show the theoretical predictions for the FFS (solid line), BG (dotted line) and RB (dashed line) methods. The symbols show the simulation results. Circles: FFS method, squares: BG method, triangles: RB method (with Metropolis acceptance/rejection). Simulation results were obtained with $400$ blocks of $N_0=1000$ starting points for FFS and $2000$ starting points per block for BG and RB. (a): ${\mathcal{C}}$ as a function of $k$, for $n=7$. (b): ${\mathcal{C}}$ as a function of $n$, for $k=3$, for evenly spaced interfaces. \label{cost_comp}}
\end{center}
\end{figure}

Figure \ref{cost_comp} shows a comparison between the simulation values of ${\mathcal{C}}$ and the theoretical predictions (Eqs (\ref{cost_ffs22}), (\ref{cost_bg}) and (\ref{cost_ros})), for the three methods, as a functions of $k$ (Fig.\ref{cost_comp}a) and of $n$ (Fig.\ref{cost_comp}b). In these calculations, the same value of $k$ was used for all interfaces: $k_i=k$ for all $i$. To obtain the data in Fig.\ref{cost_comp}b, we used interfaces which were evenly spaced in $\lambda$ and a fixed value $k=3$. We observe remarkably good agreement between the predicted and observed values for the cost, verifying that at least for this problem, Eqs  (\ref{cost_ffs22}), (\ref{cost_bg}) and (\ref{cost_ros}) are very accurate. 

\begin{figure}[h]
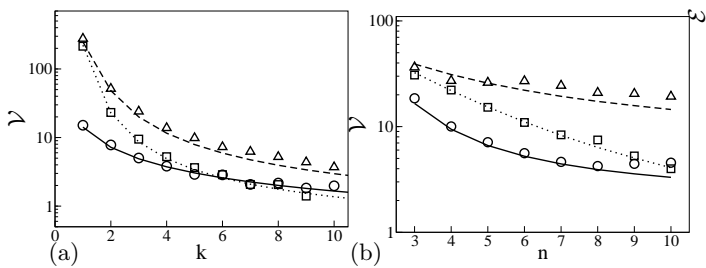

\begin{center}
\makebox[-15pt][l]{(a)}{\rotatebox{0}{{\includegraphics[scale=0.19,clip=true]{fig9a.eps}}}}\makebox[0pt][l]{(b)}{\rotatebox{0}{{\includegraphics[scale=0.19,clip=true]{fig9b.eps}}}}
\caption{Predicted and measured values of ${\mathcal{V}}$, for the Maier-Stein problem. The lines show the theoretical predictions for the FFS (solid line), BG (dotted line) and RB (dashed line) methods. The symbols show the simulation results. Circles: FFS method, squares: BG method, triangles: RB method (with Metropolis acceptance/rejection).  Simulation results were obtained with $400$ blocks of $N_0=1000$ starting points for FFS and $2000$ starting points per block for BG and RB. Interfaces were evenly spaced between $\lambda_A=-0.7$ and $\lambda_B=0.7$ (a): ${\mathcal{V}}$ as a function of $k$, for $n=7$. (b): ${\mathcal{V}}$ as a function of $n$, for $k=3$. In (b), the landscape contribution is not included in the theoretical calculation. \label{var_comp}}
\end{center}
\end{figure}

\begin{figure}[h]
\begin{center}
{\rotatebox{0}{{\includegraphics[scale=0.25,clip=true]{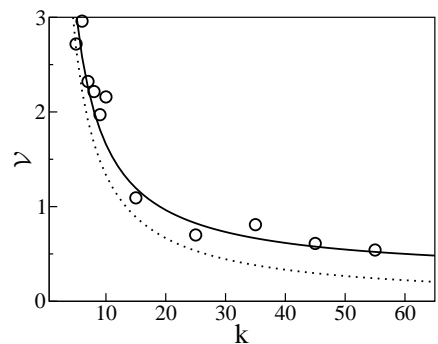}}}}
\caption{Predicted and measured values of ${\mathcal{V}}$, for the Maier-Stein problem, for the FFS method. Solid line: Eq.(\ref{ffs_imp_int}) (with landscape variance), dotted line: Eq.(\ref{ffs_imp}) (no landscape variance), circles: simulation results.\label{var_comp_ffs}}
\end{center}
\end{figure}

The predicted and measured values of ${\mathcal{V}}$ are shown in Figure \ref{var_comp}, for all three methods. Agreement is again excellent, showing that the approximations of Section \ref{err_sect} are justified, at least for this problem. The landscape contribution to ${\mathcal{V}}$ is included in Figure \ref{var_comp} for panel (a) but not for (b). In Figure \ref{var_comp_ffs}, we show the effect of neglecting this contribution (note the altered scales on both axes). Although the landscape contribution is small for this problem, it becomes significant for large $k$ as the ``binomial'' contribution decreases.

\begin{figure}[h]
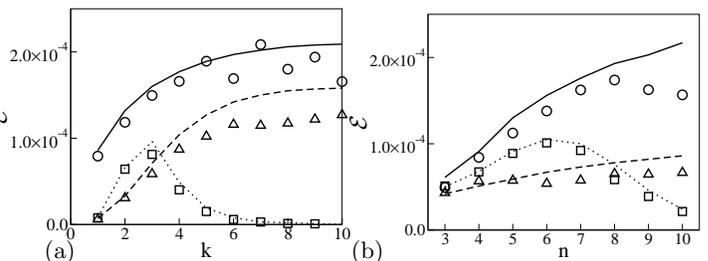

\begin{center}
\makebox[-20pt][l]{(a)}{\rotatebox{0}{{\includegraphics[scale=0.19,clip=true]{fig11a.eps}}}}\makebox[0pt][l]{(b)}{\rotatebox{0}{{\includegraphics[scale=0.19,clip=true]{fig11b.eps}}}}
\caption{Predicted and measured efficiency ${\mathcal{E}}$, for the Maier-Stein system.  The lines show the theoretical predictions for the FFS (solid line), BG (dotted line) and RB (dashed line) methods. The symbols show the simulation results. Circles: FFS method, squares: BG method, triangles: RB method (with Metropolis acceptance/rejection).  Simulation results were obtained with $400$ blocks. For FFS, each block had $N_0=1000$ starting points and for BG and RB each blocks had 2000 starting points. Interfaces were evenly spaced. (a): ${\mathcal{E}}$ vs $k$ for $n=7$. (b): ${\mathcal{E}}$ vs $n$ for $k=3$. \label{eff_comp}}
\end{center}
\end{figure}

The efficiency ${\mathcal{E}}$ is plotted in Figure \ref{eff_comp}. Excellent agreement is obtained between simulation and theory. It is also interesting to note that the trends in ${\mathcal{E}}$ as a function of $k$ are qualitatively very similar to those obtained for the model problem of Fig. \ref{eff_fig}. The BG method shows high efficiency only within a relatively narrow range of parameter values, while the FFS and RB methods are much more robust to changes in the parameters. The RB method is consistently less efficient than FFS, due to the acceptance/rejection step. As the number of interfaces $n$ becomes large, we would expect the correlations between interfaces (which are not included in our analysis) to have a greater effect, and the theoretical predictions to become less accurate. This effect is observed to a certain extent: the efficiency of FFS, for example, decreases relative to the predicted value as $n$ increases. However, this is not a dramatic effect, and in fact, even on increasing $n$ further, as far as 100 interfaces, we find a decrease of only a few percent in the efficiency of FFS. It seems therefore, that for FFS at least, one can use any number $n$ of interfaces, as long as $n$ is not too small or so very large that memory requirements become the limiting factor. 

The remarkable agreement between the theoretical predictions and the simulation results shown in Figures \ref{cost_comp}, \ref{var_comp} and \ref{eff_comp} perhaps reflects the simplicity of the Maier-Stein problem. The main assumption for the calculation of ${\mathcal{V}}$ - that the sampling of $p_i$ at different interfaces is uncorrelated - seems to be well justified in this case. We would expect our theoretical predictions to be less accurate for more complex problems, perhaps with strong correlations between interfaces. In fact, on investigating the two examples presented in our previous paper \cite{long1} - the flipping of a genetic switch and the translocation of a polymer through a pore - we find that the quantitative estimates of both the cost and variance can differ by a factor of about 10 from the theoretical predictions. Even with this caveat, however, we believe that the expressions of Section \ref{efficiency} will prove to be of practical use for a wide range of rare event simulation problems.

\section{Discussion}\label{discuss}
In this paper, we have derived simple analytical expressions for the computational cost of the three FFS-type rare event simulation methods and the statistical accuracy of the resulting estimate of the rate constant. The expressions were found to be in remarkably good agreement with simulation results for the two-dimensional non-equilibrium rare event problem proposed by Maier and Stein \cite{ms_pre93,ms_jsp96,l_prl99}. 

Our analysis allows us to draw some general conclusions about the relative merits of the three FFS-type methods. Firstly, the optimum efficiencies of the methods are all of the same order of magnitude, at least for the simple test problem studied here. However, the methods show very different sensitivities to the choice of parameters. The Branched Growth method in particular is highly sensitive, performing well only for a narrow range of parameter values. Within this range, however, it performs well in comparison to the other methods. The FFS method is the most robust to changes in the parameters, performing consistently well, even for parameter values where the other methods are very inefficient. The Rosenbluth method is lower in efficiency than the others, as a consequence of the Metropolis acceptance/rejection step which is required in order to obtain paths with the correct weights in the Transition Path Ensemble. 

These observations provide a very useful guide for choosing a rate constant calculation method. In general, unless one has a very good idea of the optimum parameters, the BG method carries a risk of being low in efficiency. Of course, strategies could be envisaged to overcome this problem - for example, one could imagine terminating a certain percentage of the branches to avoid the high cost of sampling later interfaces. The analysis used here could easily be extended to predict the likely success of such approaches. The RB method appears from this analysis to be of relatively low efficiency. However, that  is not to say that one should not use the Rosenbluth method. On the contrary, this is the only method which generates unbranched paths, making it highly suitable for situations where one wishes to analyse the paths, in order to study the transition mechanism. The RB and BG methods also require much less storage of system configurations than FFS (for which all $N_i$ points at interface $i$ must be stored in memory), making them potentially suitable for large systems. As a general conclusion, however, the results of this paper show that the FFS method is highly robust to parameter changes and is probably the method of choice for calculations of the rate constant where effects such as the storage of many configurations in memory are not important. 

These results could also suggest possible strategies for choosing the parameters for the three methods. One approach would be to use the analytical expressions derived here in an optimization scheme for finding $\{k_i\}$, $\{\lambda_i\}$ and $n$. This is likely to be useful for the BG method, but may be less essential for the FFS and RB methods, where the choice of parameters is much less critical.

We expect that the predictions of the cost and statistical error derived here will be useful not only for parameter optimization, but also for assessing, before beginning a calculation, which method to use and, indeed, whether to proceed at all. Some preliminary calculation would be  needed in order to obtain rough estimates for $R$,$S$, $P_B$, $\{p_i\}$ and (if required) $\{U_i\}$. These preliminary calculations are expected to be much cheaper than a full simulation. While the expressions for the cost and variance will be less accurate if only rough estimates for the parameters are available, we expect the results to be nevertheless accurate enough to be of use.

Furthermore, the expressions for ${\mathcal{V}}$ can be used, after  a rate constant calculation has  been completed, to obtain error bars on the calculated value of $k_{AB}$. In this case, the values of $P_B$ and $\{p_i\}$ are known. The intrinsic variances  $\{U_i\}$ can also be easily obtained during the rate constant calculation, as explained in Appendix \ref{int_var_meas}. These values  can be substituted into the expressions to obtain a reliable estimate of the statistical error in the resulting rate constant.

In this work, we provide a way to compare the efficiency of the three FFS-type methods. It would also be very useful to compare their efficiency to that of other methods, such as the method of Crooks and Chandler \cite{crooks} for non-equilibrium rare event problems, or TPS \cite{tps} or Transition Interface Sampling (TIS) \cite{vanerp,vanerp2005} for equilibrium problems. We have carried out preliminary calculations using the Crooks-Chandler method for the Maier-Stein system. We find that the value of the rate constant is in agreement with that of the FFS-type methods, but that the FFS-type methods are much more efficient. However, a thorough comparison would require a detailed investigation, optimizing the parameter choices of all the methods. We therefore leave this to a future study.

In conclusion, we have presented expressions for the computational cost and statistical accuracy of three recently introduced rare event simulation methods. We believe that the expressions presented here will be valuable in using these methods to compute rate constants and in evaluating the results of such computations.

\begin{acknowledgments}
The authors thank Axel Arnold for his careful reading of the manuscript. This work is part of the research program of the "Stichting voor Fundamenteel Onderzoek der Materie (FOM)", which is financially supported by the "Nederlandse organisatie voor Wetenschappelijk Onderzoek (NWO)''. R.J.A. was funded by the European Union Marie Curie program. 
\end{acknowledgments}

\appendix

\section{Cost of trial runs}\label{pl_app}
In order to estimate the cost of a trial run, we assume that the system undergoes one-dimensional diffusion along the $\lambda$ coordinate, with a constant drift velocity (the origin of which is a force due to the ``free energy barrier''). The problem is then equivalent to that of a particle which undergoes diffusion with drift along the $x$ axis, after being released  between two absorbing boundaries. We are interested in the mean time $\tau_\leftarrow$ or $\tau_\rightarrow$ that the particle takes to be captured at the left or right boundary, {\em{given}} that it is eventually captured at that particular boundary. Farkas and F{\"{u}}l{\"{o}}p have studied the problem of one dimensional diffusion with drift \cite{farkas}. They give analytical expressions for the probabilities $n_\leftarrow$ and $n_\rightarrow$ that the particle is absorbed at the left and right boundaries, respectively, and the rates of absorption, $j_\leftarrow$ and $j_{\rightarrow}$ at the left and right boundaries. The mean first passage time $\tau$ is the average time before the particle is absorbed at one of the boundaries:
\begin{equation}\label{tau1}
\tau = \int_0^\infty t\left[j_\leftarrow + j_\rightarrow \right] dt
\end{equation}
To compute $\tau_\leftarrow$ and $\tau_\rightarrow$, we require integrals similar to Eq.(\ref{tau1}), but including only events where the particle reaches the desired boundary. The integrals must also be normalized by the probability of reaching that boundary:
\begin{eqnarray}\label{tau2}
\tau_\leftarrow = \frac{\int_0^\infty t j_\leftarrow dt}{n_\leftarrow} \qquad ; \qquad \tau_\rightarrow = \frac{\int_0^\infty t j_\rightarrow dt}{n_\rightarrow}
\end{eqnarray}
Carrying out the integrals (\ref{tau2}) using the  expressions of  Farkas and F{\"{u}}l{\"{o}}p for $j_\leftarrow$, $j_{\rightarrow}$, $n_\leftarrow$ and $n_\rightarrow$ (Eqs (3-5) of their paper \cite{farkas}), we arrive at:
\begin{eqnarray}
\nonumber && \tau_\leftarrow = \frac{L}{v}\left[\coth{\left(\frac{Lv}{2D}\right)} - (1-\alpha) \coth{\left(\frac{(1-\alpha) Lv}{2D}\right)}\right]\\ && \tau_\rightarrow = \frac{L}{v}\left[\coth{\left(\frac{Lv}{2D}\right)} - \alpha \coth{\left(\frac{\alpha Lv}{2D}\right)}\right]
\end{eqnarray}
where $v$ is the drift velocity, $D$ is the diffusion constant, the absorbing boundaries are at $x=0$ and $x=L$ and the particle is released at $x=\alpha L$ at time $t$. In the limit that the drift velocity is large, $\cosh{\left[Lv/(2D)\right]} \to 1$ and $\tau_\leftarrow$ and $\tau_\rightarrow$ reduce to:
\begin{eqnarray}
 \tau_\leftarrow = \frac{\alpha L}{v} \qquad ; \qquad \tau_\rightarrow = \frac{(1-\alpha)L}{v}
\end{eqnarray}
In this case, the average time for a particle to be captured at a specified boundary is linearly proportional to the distance between the starting point of the particle and that boundary, and the proportionality constant is the same for particles moving against or with the drift velocity. It is therefore appropriate to approximate the cost of a trial run between $\lambda_i$ and $\lambda_j$ by $S|\lambda_j-\lambda_i|$, as in Eq.(\ref{cost1}).

\section{Acceptance probability for the RB method}\label{app_theta}
This section is concerned with the Metropolis acceptance/rejection step in the Rosenbluth method. We derive  the approximate expression (\ref{theta_1}) for the probability $\theta_i$ that a newly generated estimate $p_i^{e({\rm{n}})}=N_s^{(i)}/k_i$ for the probability $p_i$ is accepted. Upon reaching interface $i$, we calculate the Rosenbluth factor $W_i^{({\rm{n}})} = \prod_{j=0}^{i-1} N_s^{(j)}$ corresponding to the newly generated path leading to interface $i$. We compare this to the  Rosenbluth factor $W_i^{({\rm{o}})}$ corresponding to the previous path to have been accepted at interface $i$. Acceptance occurs  if the ratio $Z_i \equiv W_i^{({\rm{n}})}/W_i^{({\rm{o}})}$  is greater than a random number $0 < s < 1$. If we know the distribution function $P(Z_i)$, the acceptance probability is given by:
\begin{eqnarray}\label{accc1}
\theta_i &=& \int_0^1 ds \,\, \int_s^\infty dZ_i\,\, P(Z_i)
\end{eqnarray}
We would therefore like to calculate $P(Z_i)\equiv P(W_i^{({\rm{n}})}/W_i^{({\rm{o}})})$.  To obtain this, we require the distribution functions for both $W_i^{({\rm{n}})}$ and $W_i^{({\rm{o}})}$. We begin with $W_i^{({\rm{n}})}$, which we can write as
\begin{equation}\label{logw}
\log{[W_i^{({\rm{n}})}]} = \sum_{j=0}^{i-1} \log{[N_s^{(j)}]}
\end{equation}
We assume that the $\log{[N_s^{(j)}]}$ for each interface $j$ are independent variables ({\em{i.e.}} that the sampling at different interfaces is uncorrelated). Since we are adding many independent variables, we  apply the Central  Limit Theorem \cite{riley} to Eq.(\ref{logw}). In the limit of a large number of interfaces, the distribution of $y_i^{({\rm{n}})}=\log{[W_i^{(\rm{n})}]}$, is:
\begin{equation}\label{py}
p(y_i^{({\rm{n}})}) = \frac{1}{\sigma_i\sqrt{2\pi}}\exp{\left[-\frac{(y_i^{({\rm{n}})}-\mu_i)^2}{2\sigma_i^2}\right]}
\end{equation}
where
\begin{equation}\label{muu}
\mu_i = \sum_{j=0}^{i-1} E[\log{N_s^{(j)}}]
\end{equation}
and 
\begin{equation}\label{sigg}
\sigma_i^2 = \sum_{j=0}^{i-1} V[\log{N_s^{(j)}}]
\end{equation}
The expectation value $E[\log{N_s^{(j)}}]$  can be found approximately by performing a Taylor expansion of $\log{N_s^{(j)}}$ about $E[N_s^{(j)}]$, to give:
\begin{eqnarray}\label{Elog1}
\log{N_s^{(j)}} &\approx& \log{E[N_s^{(j)}]}+\frac{\left(N_s^{(j)}-E[N_s^{(j)}]\right)}{E[N_s^{(j)}]} \\\nonumber && - \frac{1}{2}\frac{\left(N_s^{(j)}-E[N_s^{(j)}]\right)^2}{E[N_s^{(j)}]^2}
\end{eqnarray}
taking the expectation value of Eq.(\ref{Elog1}), we obtain:
\begin{eqnarray}\label{Elog2}
E[\log{N_s^{(j)}}] &\approx& \log{E[N_s^{(j)}]}-\frac{V[N_s^{(j)}]}{2E[N_s^{(j)}]^2}
\end{eqnarray}
Using the variance relation (\ref{veq2}), we find that 
\begin{equation}\label{Vlog2}
V[\log{N_s^{(j)}}] \approx \frac{1}{E[N_s^{(j)}]^2}V[N_s^{(j)}]
\end{equation}
We now need to know $E[N_s^{(j)}]$ and $V[N_s^{(j)}]$. On firing $k_i$ trials from interface $i$, we know that the number of successes follows a binomial distribution. However, the variable $N_s^{(j)}$ in Eqs (\ref{Elog}) and (\ref{Vlog}) refers to the number of successes at interface $j$, given that we know the path subsequently reached interface $i>j$. We therefore know that $N_s^{(j)}>0$, so that
\begin{equation}
p(N_s^{(j)}) = \frac{1}{(1-q_j^{k_j})}\frac{k_j!}{(k_j-N_s^{(j)})!(N_s^{(j)})!}p_j^{N_s^{(j)}}q_j^{k_j-N_s^{(j)}}
\end{equation}
so that 
\begin{equation}\label{Ns1}
E(N_s^{(j)}) = \frac{k_jp_j}{(1-q_j^{k_j})}
\end{equation}
\begin{equation}
E({N_s^{(j)}}^2) = \frac{\left[k_jp_jq_j+k_j^2p_j^2\right]}{(1-q_j^{k_j})}
\end{equation}
and
\begin{equation}\label{Ns2}
V[N_s^{(j)}] = \frac{\left[(1-q_j^{k_j})k_jp_jq_j - k_j^2p_j^2q_j^{k_j}\right]}{(1-q_j^{k_j})^2}
\end{equation}
Substituting (\ref{Ns1}) and (\ref{Ns2}) into (\ref{Elog2}) and (\ref{Vlog2}), we obtain:
\begin{eqnarray}\label{Elog}
E[\log{N_s^{(j)}}] &\approx&  \log{\left[\frac{k_jp_j}{1-q_j^{k_j}}\right]} - \frac{1}{2}\left[\frac{(1-q_j^{k_j})q_j}{k_jp_j}-q_j{^k_j}\right]
\end{eqnarray}
and
\begin{equation}\label{Vlog}
V[\log{N_s^{(j)}}] \approx  \frac{q_j(1-q_j^{k_j})}{k_jp_j} -q_j^{k_j}
\end{equation}
Substituting (\ref{Elog}) and (\ref{Vlog}) in turn into (\ref{muu}) and (\ref{sigg}) leads to
\begin{equation}
\mu_i = \sum_{j=0}^{i-1} \log{\left[\frac{k_jp_j}{(1-q_j^{k_j})}\right]} - \frac{1}{2}\left[\frac{(1-q_j^{k_j})q_j}{k_jp_j}-q_j^{k_j}\right]
\end{equation}
and 
\begin{equation}
\sigma_i^2 = \sum_{j=0}^{i-1} \frac{q_j(1-q_j^{k_j})}{k_jp_j} -q_j^{k_j} 
\end{equation}
Finally, the distribution function $f(W_i)$ for the Rosenbluth factor of the newly generated path can be found by making the change of variables $W_i = \exp{[y_i^{(\rm{n})}]}$ in Eq.(\ref{py}), to give:
\begin{equation}\label{new1}
f(W_i) = \frac{1}{\sigma_i\sqrt{2\pi}}\left[\frac{1}{W_i}\right]\exp{\left[-\frac{(\log{[W_i]}-\mu_i)^2}{2\sigma_i^2}\right]}
\end{equation}
We now turn to the distribution function $g(W_i)$ for the Rosenbluth factor $W_i^{({\rm{o}})}$ of the previous path to have been accepted at interface $i$. $W_i^{({\rm{o}})}$ does not follow the same distribution as $W_i^{({\rm{n}})}$, because  the ``previous'' path has survived at least one round of acceptance/rejection. We know that the acceptance/rejection procedure re-weights paths by  a factor proportional to the Rosenbluth factor (see Section {\ref{sec_ros}}), so if we assume that $W_i^{({\rm{o}})}$ has been ``fully'' re-weighted (note that this is an approximation), we can say that 
\begin{equation}\label{fw}
g(W_i) \approx \frac{W_if(W_i}{\int_{0}^{\infty}W' f(W') dW'}
\end{equation}
The denominator of Eq.(\ref{fw}) ensures that $g(W_i)$ is properly normalized. Substituting (\ref{fw}) into  (\ref{new1}), we find that:
\begin{equation}\label{old1}
g(W_i) = \frac{1}{I} \frac{1}{\sqrt{2\pi}\sigma_i}\exp{\left[-\frac{(\log{[W_i]}-\mu_i)^2}{2\sigma_i^2}\right]}
\end{equation}
where
\begin{equation}
I=\int_{0}^{\infty}W_i f(W_i) dW_i =  \exp{\left[\mu_i + \frac{\sigma_i^2}{2}\right]}
\end{equation}
Armed with Eqs (\ref{new1}) and (\ref{old1}), we can now find the distribution function $P(Z_i)$ for the ratio $Z_i \equiv W_i^{({\rm{n}})}/W_i^{({\rm{o}})}$. This is given by:
\begin{equation}
P(Z_i) = \int_{0}^{\infty}\int_{0}^{\infty}dW_i dW_i' g(W_i) \, f(W_i') \, \delta \left(\frac{W_i'}{W_i}-Z_i\right)
\end{equation} 
Changing the variable of the second integral to $Z_i'=W_i'/W_i$, we obtain
\begin{eqnarray}\label{one}
\nonumber P(Z_i) &=& \int_{0}^{\infty}\int_{0}^{\infty}dW_i dZ_i'\,\, W_i g(W_i)\, f(Z_i' W_i)\, \delta \left(Z_i'-Z_i\right)\\ &=& \int_{0}^{\infty}dW_i\,\, W_i\, g(W_i) \, f(Z_i W_i) 
\end{eqnarray}
Substituting (\ref{new1}) and (\ref{old1}) into (\ref{one}), we obtain:
\begin{eqnarray}
P\left(Z_i\right) && = \frac{1}{2\pi\sigma_i^2 I Z_i} \times \\\nonumber && \hspace{-0.3cm} \int_{0}^{\infty}dW_i\,\,\exp{\left[-\frac{(\log{[W_i]}-\mu_i)^2 + (\log{[Z_iW_i]}-\mu_i)^2}{2\sigma_i^2}\right]}
\end{eqnarray}
This integral can be carried out analytically \cite{gradshteyn}, to give:
\begin{eqnarray}\label{pz}
P\left(Z_i\right)  = \frac{\exp{\left[-\frac{\sigma_i^2}{4}\right]}}{2\sigma_i Z_i \sqrt{\pi}} \exp{\left[-\frac{(\log{Z_i})^2}{4\sigma_i^2}-\frac{\log{Z_i}}{2}\right]}
\end{eqnarray}
We are now finally in a position to calculate the acceptance probability $\theta_i$, using Eq.(\ref{accc1}). Substituting Eq.(\ref{pz}) into (\ref{accc1}) and integrating over $Z_i$, we obtain  \cite{abramowitz}:
\begin{eqnarray}\label{accc2}
\theta_i &=&  \frac{1}{2}\int_0^1 ds \,\, \left[1-\frac{\sqrt{\pi}}{2}{\rm{Erf}}\left[\frac{\sigma_i}{2}+\frac{\log{s}}{2\sigma_i}\right]\right]\\\nonumber &=&\frac{1}{2}-\frac{\sqrt{\pi}}{4}\left[2{\rm{Erf}}\left(\frac{\sigma_i}{2}\right)-1\right]
\end{eqnarray}
where ${\rm{Erf}}(x)$ is the error function: ${\rm{Erf}}(x)=(2/\sqrt{\pi})\int_0^x e^{-t^2}dt$. 

\begin{figure}[h]
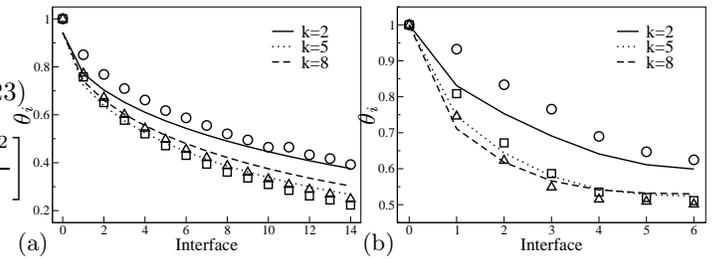

\begin{center}
\makebox[0pt][l]{(a)}{\rotatebox{0}{{\includegraphics[scale=0.19,clip=true]{fig12a.eps}}}}\makebox[0pt][l]{(b)}{\rotatebox{0}{{\includegraphics[scale=0.19,clip=true]{fig12b.eps}}}}
\caption{(a): ``Simulated'' and predicted acceptance probabilities $\theta_i$ for interfaces $0 \le i \le 14$, for the ``simulated simulation'' described in the text, for $k=2,5,8$. (b): Simulated and predicted values of $\theta_i$ for $0 \le i \le 6$ for the Maier-Stein problem of Section \ref{sec_ms}, for $k=2,5,8$. In both plots, solid lines represent predicted values for $k=2$, dotted lines, $k=5$ and dashed lines, $k=8$. Symbols represent simulation results: circles: $k=2$, squares: $k=5$ and triangles: $k=8$.\label{acc_fig} }
\end{center}
\end{figure}

Although Eq.(\ref{accc2}) is a simple and convenient expression for the acceptance probability $\theta_i$, its derivation required several approximations. We have therefore tested the validity of Eq.(\ref{accc2}). We first carried out a ``simulated simulation'', in which we defined a series of $N=15$ interfaces, each with the same value of $p_i=p=10^{-6/15}$, and ``simulated'' the Rosenbluth calculation, each time drawing a random number to determine the outcome of a given ``trial run'', for a given number of trial runs $k_i=k$, taken to be the same for all interfaces. We measured the acceptance probabilities at each interface after $2 \times 10^6$ Rosenbluth ``path generations'', and compared these to Eq.(\ref{accc2}). The results are shown in Figure \ref{acc_fig}a, for $k=2$, $k=5$ and $k=8$. The agreement with the ``simulation'' is very reasonable. To compare with real simulation results, we also measured the acceptance probabilities $\theta_i$ for the RB simulations of the Maier-Stein system described in Section \ref{sec_ms}. The results are compared with the predictions of Eq.(\ref{accc2}) in Figure \ref{acc_fig}b. Again, quite good agreement is obtained. 

\section{The effects of landscape variance}\label{int_var_inc}
In this section, we include the effects of the ``landscape variance'' in our expressions for the relative variance ${\mathcal{V}}$ of $P_B^e$.  The result will be that expressions  (\ref{ffs_imp}), (\ref{bg_imp}) and (\ref{rb_imp}) are transformed into  (\ref{ffs_imp_int}), (\ref{bg_imp_int}) and (\ref{rb_imp_int}). As described in Section \ref{efficiency}, we suppose that point $j$ at interface $\lambda_i$ has probability $p_i^{(j)}$ that a trial run fired from it will reach $\lambda_{i+1}$, rather than $\lambda_A$. The variance in the $p_i^{(j)}$ values for points at $\lambda_i$ (sampled according to their expected occurrence in the  trial run firing procedure) is the ``landscape variance'', $U_i$.

If we choose a particular point $j$, fire $k_i$ trial runs and measure the number of successes $N_s^{(i)}$, we expect to obtain a mean value $E[N_s^{(i)}|j]=k_ip_i^{(j)}$, and a  variance $V[N_s^{(i)}|j]=k_ip_i^{(j)}q_i^{(j)}$ (in analogy with Eqs (\ref{simp_e}) and (\ref{simp})).  We now average over many points $j$ at interface $\lambda_i$, using the general variance relation (\ref{veq1}):
\begin{eqnarray}
V[N_s^{(i)}] &=& E\left[V\left[N_s^{(i)}|j\right]\right] + V\left[E\left[N_s^{(i)}|j\right]\right]\\\nonumber &=& E\left[k_ip_i^{(j)}q_i^{(j)}\right] + V\left[k_ip_i^{(j)}\right]
\end{eqnarray}
where the mean and the variance are taken over the distribution of points $j$. Since $E[p_i^{(j)}q_i^{(j)}] = E[p_i^{(j)}-(p_i^{(j)})^2]= E[p_i^{(j)}]-E[(p_i^{(j)})^2]$ and  $U_i=E[({p_i^{(j)}})^2]-(E[p_i^{(j)}])^2$, we can deduce that $E\left[k_ip_i^{(j)}q_i^{(j)}\right]=k_i(p_i-p_i^2-U_i)=k_i(p_iq_i-U_i)$. Using Eq.(\ref{veq3}), we have $V[k_ip_i^{(j)}]=k_i^2V[p_i^{(j)}]=k_i^2U_i$, so that
\begin{eqnarray}\label{intvar}
V[N_s^{(i)}] = k_ip_iq_i + U_ik_i^2\left[1-\frac{1}{k_i}\right]
\end{eqnarray}
This first term on the r.h.s. of Eq.(\ref{intvar}) corresponds to Eq.(\ref{simp}): the binomial contribution arising from the limited number of trial runs per point. The second term is an extra contribution, due to the landscape variance.

We now repeat the derivations of Section \ref{pb_sec}, simply replacing Eq.(\ref{simp}) by Eq.(\ref{intvar}). We begin with the RB method, for which Eq.(\ref{ros111}) becomes
\begin{eqnarray}
V^{\rm{rb}}[p_i^{e}] &=& \left[\frac{1}{N_0}\right]\left[\frac{p_iq_i}{k_i} + U_i\left(1-\frac{1}{k_i}\right)\right]\\\nonumber && \times \left[\frac{(2-\theta_i)}{\theta_i}\right]\frac{\left(1-q_i^{k_i}\right)}{ \prod_{j=0}^{i}\left(1-q_j^{k_j}\right)}
\end{eqnarray}
and Eq.(\ref{rb_imp}) is replaced by Eq.(\ref{rb_imp_int}):
\begin{eqnarray}
\nonumber {\mathcal{V}}^{\rm{rb}} =  \sum_{i=0}^{n-1} && \Bigg\{\left[\frac{q_i}{p_ik_i} + \frac{U_i}{p_i^2}\left(1-\frac{1}{k_i}\right)\right]\\\nonumber && \times \left[\frac{(2-\theta_i)}{\theta_i}\right]\frac{\left(1-q_i^{k_i}\right)}{ \prod_{j=0}^{i}\left(1-q_j^{k_j}\right)}\Bigg\}
\end{eqnarray}
For the BG method, Eq.(\ref{bgvar2}) is replaced by:
\begin{eqnarray}
V[N_s^{(i)}] &= \left[k_i p_i q_i+ U_i \left(k_i^2-k_i\right)\right]\prod_{j=0}^{i-1}k_jp_j \\\nonumber &+ k_i^2 p_i^2 V[N_s^{(i-1)}]&\qquad (i>0)\\\nonumber &= k_i p_i q_i+ U_i \left(k_i^2-k_i\right) &\qquad (i=0)
\end{eqnarray}
and Eq.(\ref{bg_imp}) becomes Eq.(\ref{bg_imp_int}):
\begin{equation}
\nonumber {\mathcal{V}}^{\rm{bg}} = \sum_{i=0}^{n-1} \left[\frac{k_iq_ip_i + U_i \left(k_i^2-k_i\right)}{k_ip_i\prod_{j=0}^{i}p_j k_j}\right]
\end{equation}

For the FFS method, the situation is slightly more complicated, because the number of trials fired from point $j$ at interface $i$ is not fixed. We make $M_i$ trials from the $N_i$ points at $\lambda_i$, each time selecting a starting point at random (so that the probability a particular point is chosen is $1/N_i$). Since we no longer assume that all points at interface $i$ are identical, we must now take account of the distribution of the number of times $m_j$  that point $j$ is selected. This is in fact  a multinomial distribution \cite{riley,wolfram}, which has average $E[m_j]=M_i/N_i$ and variance $V[m_j]=M_i\left[1/N_i (1-1/N_i)\right]$. Let us now do a ``thought experiment'' in which we first decide how many trial will be fired from each point - {\em{i.e.}} we fix the set of values $\{m_j\}$ (of course, $\sum_j m_j = M_i$). We then fire these trials and measure the total number $N_s^{tot}$ which reach $\lambda_{i+1}$. The expectation value for $N_s^{tot}$ is  
\begin{equation}\label{111}
E[N_s^{tot}|\{m_j\}] = \sum_j m_jp_i^{j} = M_ip_i
\end{equation} 
and the variance is found using Eq.(\ref{intvar}), with $k_i$ replaced by $m_j$, multiplying by $m_j^2$ and summing over all $j$:
\begin{equation}\label{222}
V[N_s^{tot}|\{m_j\}] = \sum_j \left[m_jp_iq_i + U_i\left[m_j^2-m_j\right]\right]
\end{equation}
We now imagine that we average the results over many sets of values $\{m_j\}$.  Using the general relation (\ref{veq1}), we obtain:
\begin{eqnarray}\label{vns1}
V[N_s^{tot}] &=& V\left[E[N_s^{tot}|\{m_j\}]\right] + E\left[V[N_s^{tot}|\{m_j\}]\right] \\\nonumber &=& V\left[M_ip_i\right] + E\left[M_ip_iq_i + U_i\sum_j m_j^2 -U_iM_i\right] \\\nonumber &=& M_ip_iq_i + U_i\left[N_iE[m_j^2]-M_i\right]
\end{eqnarray}
Here, the variance and expectation are with respect to the distribution of $\{m_j\}$ values. The last line follows from the fact that $V\left[M_ip_i\right]=0$ as both $M_i$ and $p_i$ are constants with respect to changes in $\{m_j\}$. Since  $V[m_j]=M_i\left[1/N_i (1-1/N_i)\right]=E[m_j^2]-E[m_j]^2$, we find that $E[m_j^2]=(M_i/N_i)(1-1/N_i) + M_i^2/N_i^2$. Substituting this into Eq.(\ref{vns1}), we obtain:
\begin{eqnarray}\label{vns2}
V[N_s^{tot}] = M_ip_iq_i + \frac{U_i}{N_i}\left[M_i^2 - M_i]\right]
\end{eqnarray}
Since $p_i^e=N_s^{tot}/M_i$, we must divide Eq.(\ref{vns2}) by $M_i^2$ to obtain $V[p_{i}^{e}]^{\rm{ffs}}$:
\begin{equation}
V[p_{i}^{e}]^{\rm{ffs}} = \frac{p_iq_i}{M_i} + \frac{U_i}{N_i}\left(1-\frac{1}{M_i}\right)
\end{equation}
This leads to:
\begin{eqnarray}
\nonumber {\mathcal{V}}^{\rm{ffs}} =  N_0\sum_{i=0}^{n-1} && \Bigg\{\left[\frac{q_i}{p_i M_i} + \frac{U_i}{p_i^2 N_i}\left(1-\frac{1}{M_i}\right)\right]\\ && \times\frac{\left(1-q_i^{M_i}\right)}{ \prod_{j=0}^{i}\left(1-q_j^{M_j}\right)}\Bigg\}
\end{eqnarray}
where $N_i=M_{i-1}p_{i-1}$ for $i>0$ and $N_i=N_0$ for $i=0$. Rewriting in terms of $k_i \equiv M_i/N_0$, we obtain Eq.(\ref{ffs_imp_int}):
\begin{eqnarray}
\nonumber  {\mathcal{V}}^{\rm{ffs}} = \sum_{i=0}^{n-1} && \Bigg\{\left[\frac{q_i}{p_i k_i} + \frac{U_iN_0}{p_i^2 N_i}\left(1-\frac{1}{N_0 k_i}\right)\right]\\\nonumber && \times\frac{\left(1-q_i^{N_0k_i}\right)}{ \prod_{j=0}^{i}\left(1-q_j^{N_0k_j}\right)}\Bigg\}
\end{eqnarray}

\section{Measuring the Intrinsic Variance}\label{int_var_meas}

\begin{figure}[h]
\begin{center}
{\rotatebox{0}{{\includegraphics[scale=0.25,clip=true]{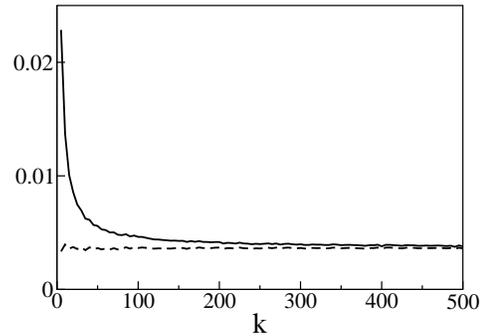}}}}
\caption{$V[N_s^{(0)}]/k^2$ (solid line) and $(1/(k-1))\left[V[N_s^{(0)}]/k-p_0q_0\right]$ (dashed line), as functions of $k=M_0/N_0$, calculated using FFS as described in Section \ref{int_var_meas}, for the Maier-Stein problem of Section \ref{sec_ms} with $10000$ points at the first interface $\lambda_0=-0.7$.\label{lam_1_Vint}}
\end{center}
\end{figure}

In this section, we describe a simple and computationally cheap procedure for measuring the landscape variance parameters $U_i$.  Given a correctly weighted collection of $N_i$ points at interface $\lambda_i$ (obtained, for example, using FFS), we could fire  an extremely large number $k$ of trial runs from each point and measure the variance among points  in the values of $N_s^{(i,j)}$ - where $N_s^{(i,j)}$ denotes the number of successful trials from point $j$:
\begin{equation}\label{meas_var1}
U_i = V[p_i]= \frac{V[N_s^{(i)}]}{k^2} = \frac{1}{k^2}\left\{\sum_{j=1}^{N_i} \frac{{N_s^{(i,j)}}^2}{N_i} - \left[\sum_{j=1}^{N_i} \frac{N_s^{(i,j)}}{N_i}\right]^2\right\} \qquad (k \to \infty)
\end{equation} 
This is likely to be an expensive procedure. Fortunately, however, it is not necessary  to fire a very large number of trial runs from each point. Instead, we make use of  expression (\ref{intvar}), which can be written as
\begin{equation}\label{meas_var2}
U_i = \frac{kV[p_i^e]-p_iq_i}{k-1} = \frac{1}{(k-1)}\left[\frac{V[N_s^{(i)}]}{k}-p_iq_i\right]
\end{equation}
where the expression now holds for any value of $k$. In the limit that $k \to \infty$, Eq.(\ref{meas_var2}) reduces to (\ref{meas_var1}). As a practical procedure, therefore, we  generate a collection of $N_i$ points at interface $i$ (using, for example, FFS), and fire $k$ trials from each point - $k$ does not have to be a large number. For each point $j$, we record the number of successful trials  $N_s^{(i,j)}$. The variance $V[N_s^{(i)}]$ of these values is inserted into Eq.(\ref{meas_var2}) to give a value for $U_i$. Figure \ref{lam_1_Vint} shows the results of this procedure for the Maier-Stein problem of Section \ref{sec_ms}. For the first interface ($\lambda_0=-0.7$), $U_i$ was calculated using Eq.(\ref{meas_var2}), using $k$ trials for each of $10000$ points collected at $\lambda_0$. The solid line is the measured value of $V[N_s^{(i)}]/k^2$, while the dashed line is the value of $U_i$ obtained from Eq.(\ref{meas_var2}). The two lines converge, of course, for large values of $k$. Figure \ref{lam_1_Vint} shows that accurate results for $U_i$ can be obtained using Eq.(\ref{meas_var2}), using only a small number of trial runs per point.


\end{document}